\documentclass[11pt, a4paper]{article}

\usepackage{jheppub}
\usepackage{amsmath, amsfonts, amsthm, amssymb, graphicx}
\usepackage[normalem]{ulem}
\usepackage{color}
\usepackage{cancel}
\usepackage{float}
\usepackage[utf8]{inputenc}
\usepackage{lipsum}
\usepackage{centernot}
\usepackage{ulem}
\usepackage{comment}
\usepackage{mathtools}
\usepackage{enumitem}
\usepackage{feynmp}
\usepackage{comment}
\newcommand{\tmax}{\text{max}}

\newcommand\figref{Fig.~\ref}

\def\be{\begin{equation}}
\def\ee{\end{equation}}
\def\ben{\begin{equation*}}
\def\een{\end{equation*}}

\def\mc#1{{\mathcal{#1}}}

\def\la{\langle}
\def\ra{\rangle}

\def\vo{\mathcal{V}}
 at 10truept

\DeclareMathAlphabet\mathbfcal{OMS}{cmsy}{b}{n}

\title{Quintessence and the Swampland: \\
\Large{The numerically controlled regime of moduli space}}

\author[a,b]{Michele Cicoli,} 
\author[c,d]{Francesc Cunillera,} 
\author[c,d]{Antonio Padilla,} 
\author[a,b]{Francisco G. Pedro} 

\affiliation[a]{Dipartimento di Fisica e Astronomia, Università di Bologna, via Irnerio 46, 40126 Bologna, Italy} 
\affiliation[b]{INFN, Sezione di Bologna, viale Berti Pichat 6/2, 40127 Bologna, Italy} 
\affiliation[c]{School of Physics and Astronomy, University of Nottingham, Nottingham NG7 2RD, UK} 
\affiliation[d]{Nottingham Centre of Gravity, University of Nottingham, Nottingham NG7 2RD, UK}

\emailAdd{michele.cicoli@unibo.it}
\emailAdd{francesc.cunilleragarcia@nottingham.ac.uk}
\emailAdd{antonio.padilla@nottingham.ac.uk}
\emailAdd{francisco.soares@unibo.it}

\abstract{We provide a detailed discussion of the main theoretical and phenomenological challenges of quintessence model building in any numerically controlled regime of the moduli space of string theory. We argue that a working quintessence model requires a leading order non-supersymmetric (near) Minkowski vacuum with an axionic flat direction. This axion, when lifted by subdominant non-perturbative effects, could drive hilltop quintessence only for highly tuned initial conditions and a very low inflationary scale. Our analysis has two important implications. Firstly, scenarios which are in agreement with the swampland conjectures, such as those that include runaways, or supersymmetric AdS and Minkowski vacua, cannot give rise to phenomenologically viable quintessence with full computational control. This raises doubts on the validity of the swampland dS conjecture since it would imply a strong tension between quantum gravity and observations. Secondly, if data should prefer dynamical dark energy, axion models based on alignment mechanisms look more promising than highly contrived hilltop scenarios.}

\begin{document}

\maketitle

\section{Introduction}

A wealth of cosmological probes, from measurements of the cosmic microwave background radiation \cite{Planck:2018vyg} to observations of distant supernova \cite{Perlmutter:1998np,SupernovaSearchTeam:1998fmf}, point to a universe that underwent two phases of accelerated expansion. The first of these occurred very early on and is often described by inflation, with  a scalar field in slow roll along a flat potential. The second phase of acceleration - dark energy - is on-going. The simplest empirical model of dark energy assumes it is driven by a cosmological constant, or vacuum energy, although the scale of the observed vacuum energy is more than 120 orders of magnitude less than expected from naturalness considerations \cite{Weinberg:1988cp,Polchinski:2006gy,Burgess:2013ara, Padilla:2015aaa}. Alternatively, dark energy could also be driven by a quintessence field  \cite{Peebles:1987ek,Ratra:1987rm,Caldwell:1997ii}, a scalar field in slow roll similar to inflation, albeit at a much lower scale \cite{Copeland:2006wr}.  Even for quintessence, we still face the question of why the vacuum energy is small and does not dominate the dynamical potential for dark energy. 

Establishing the microscopic origin of both inflation and dark energy is an important challenge for string theory phenomenology. In \cite{Obied:2018sgi} it has been conjectured that scalar potentials that can be derived from putative quantum gravity theories obey the bound
\be
V_\phi\ge \frac{c}{M_p}\,V\,,
\ee
where $c$ is a positive and dimensionless order one constant. If true, this conjecture has serious implications for inflation in the early universe and dark energy at present times. The most obvious consequence is that de Sitter (dS) vacua are forbidden, ruling out the cosmological constant as the source of dark energy. However, the bound is also in some tension with the requirement of slow roll in two derivative scalar actions, both for inflation and dark energy. While this tension is stronger in the context of inflation, it may be acceptable for dark energy models given that current bounds on $\omega_{\rm DE}$ \cite{Planck:2018vyg} are more relaxed that those derived from the scalar spectral tilt, $n_s$, for inflationary models \cite{Planck:2018jri,Amendola:2016saw}. It was later realised that this bound would rule out the experimentally tested Higgs potential, and would preclude electroweak symmetry breaking which requires $V_\phi=0$ for $V_{\phi\phi}<0$ and $V>0$ \cite{Denef:2018etk}. Moreover, it would also rule out supersymmetric AdS vacua that are accompanied by dS maxima at large field values \cite{Conlon:2018eyr}. This unsatisfactory state of affairs prompted the proposal of a refined conjecture that took the form \cite{Garg:2018reu,Ooguri:2018wrx}
\be
V_\phi\ge \frac{c}{M_p}\,V \qquad\text{or}\qquad V_{\phi\phi}\leq -\frac{c'}{M_p^2}\,V\,, 
\label{dscon}
\ee
where $c$ and $c'$ are positive and dimensionless order one constants. These conjectures are not based on rigorous proofs and several counterexamples have been proposed \cite{Louis:2012nb, Cicoli:2012vw, Cicoli:2012fh, Cicoli:2013cha, Blaback:2013ht, Braun:2015pza, Cribiori:2019hrb, Antoniadis:2019rkh, Crino:2020qwk, Cicoli:2021dhg}. Rather, the logic behind their formulation is the theoretical difficulty in  establishing the existence of a dS vacuum in a fully convincing manner, mainly due to the need to break supersymmetry. Strong evidence in favour of the refined version of the conjecture has been given in \cite{Ooguri:2018wrx} for any parametrically controlled regime of string theory using a combination of the distance conjecture and entropy considerations. This is the regime where the semiclassical approximation can be made arbitrarily good by sending  the parameters that control the string loop and the $\alpha'$ expansions to zero. These are, respectively, the real part of the axio-dilaton $S$, which sets the string coupling $g_s = 1/{\rm Re}(S)$, and the extra-dimensional volume in string units $\vo$ which controls the $\alpha'$ expansion since $1/\vo^{1/3} = \alpha' /{\rm Vol}^{1/3}$ (where ${\rm Vol}$ is the dimensionful volume). The asymptotic limit where ${\rm Re}(S)\to \infty$ and $\vo\to \infty$ corresponds to the semiclassical approximation with no dS vacua. 

However, dS vacua could still exist in the bulk of moduli space where the quality of the approximations should be carefully checked. In particular, a necessary condition to have control over the effective field theory is the existence of small expansion parameters such as the flux-generated superpotential $W_0\ll 1$ in KKLT models \cite{Kachru:2003aw} and the inverse of the internal volume $1/\vo\ll 1$ in LVS vacua \cite{Balasubramanian:2005zx, Conlon:2005ki, Cicoli:2008va}. Much progress has been made in this direction by determining perturbative \cite{Becker:2002nn, Berg:2005ja, vonGersdorff:2005ce, Berg:2007wt, Cicoli:2007xp, Ciupke:2015msa, Antoniadis:2018hqy} and non-perturbative corrections \cite{Blumenhagen:2009qh}, or by estimating their moduli dependence using higher dimensional arguments based on symmetries \cite{Burgess:2020qsc} and geometry \cite{Cicoli:2021rub}. However, it is fair to say that the existence of dS vacua in the interior of the moduli space has still to be established in a fully convincing manner and there are a growing number of no-go theorems explicitly demonstrating their absence in particular compactifications of string inspired effective theories \cite{Maldacena:2000mw, Green:2011cn, Gautason:2012tb,Dasgupta:2014pma,Kutasov:2015eba, Quigley:2015jia, Dine:2020vmr, Montero:2020rpl,Cunillera:2021fbc}. Even if they \emph{did} exist, dS vacua in string theory might well be short-lived, as suggested by the TCC conjecture \cite{Bedroya:2019snp}. Of course, this is not a problem for dark energy as observations only require it to be dominant for a single efolding of accelerated expansion. This may even be desirable in the context of the cosmological coincidence problem \cite{whynow, Zlatev:1998tr, Velten:2014nra, Cunillera:2021izz}.

All these considerations show that the existence of dS vacua in string theory is still an open problem which requires further scrutiny. It is, therefore, interesting to investigate if the alternative to a cosmological constant --- namely, quintessence --- shares the same technical difficulties. To this end, we shall focus on the microscopic origin of dark energy as a dynamically evolving scalar field emerging from a compactification of string theory.

In our companion paper \cite{us}, we have shown  that quintessence cannot be realised in any parametrically controlled regime of string theory since the dilaton and volume mode runaways in the asymptotic region of moduli space are too steep to drive an epoch of accelerated expansion. The obstruction echoes some of the obstructions to dS vacua \cite{Ooguri:2018wrx}, with related results for quintessence also being derived in  \cite{Hertzberg:2007wc, Ibe:2018ffn, Garg:2018zdg, ValeixoBento:2020ujr, Andriot:2020vlg}. Note that the situation does not improve if one performs a multifield evolution including their corresponding axionic fields. In fact, even if non-geodesic trajectories on curved field manifolds could, in principle, yield a period of accelerated expansion for steep potentials \cite{Cicoli:2020cfj, Cicoli:2020noz}, this is never the case for either ${\rm Re}(S)$ or $\vo$ \cite{BCDP}. 

As a result, quintessence can only be realised in the bulk of moduli space where it generically shares the same control issues as dS model building \cite{Cicoli:2018kdo}. On top of the technical difficulties in trusting the effective field theory, quintessence is known to feature some phenomenological challenges including the `light volume problem' and the `F-term problem' \cite{Hebecker:2019csg}. The `light volume problem' relates to quintessence driven by a saxion, typically a volume modulus. To be compatible with the acceleration we see today, this modulus needs to be extremely light, with its mass bounded above by the current Hubble scale. As it also couples to matter with gravitational strength, this would yield an additional long range  scalar force, in violation of fifth force constraints \cite{Will:2014kxa}. The `F-term problem' is  associated with radiative corrections involving supersymmetric particles running in loops, producing contributions to the scalar potential that are much larger than the dark energy scale. Traditional quintessence, at least in  a perturbative regime,  also has some observational problems, having been shown to enhance the so-called Hubble tension \cite{Colgain:2019joh,Banerjee:2020xcn} which is already at $5\sigma$ for $\Lambda$CDM \cite{Riess:2021jrx}. 

Here we add to the challenges facing quintessence in string theory. In particular, we show how a version of the so called `Kallosh-Linde (KL) problem' \cite{Kallosh:2004yh} drastically constrains the spectrum of possibilities. The KL problem is one of runaway behaviour in the volume mode during inflation. It is normally used to constrain the scale of inflation against the gravitino mass. We use it to constrain the form of the underlying scalar potential responsible for dark energy, exploiting the huge hierarchy of scales between the acceleration today and in the early universe. This hierarchy makes it extremely difficult to have a scalar potential that is compatible with current observations and is protected from the KL runaway during inflation. 

Let us briefly run through the logic. We begin with $V_0(\vo)$, the potential that fixes the volume mode. However, the volume mode also couples to any source of energy-momentum thanks to the Weyl rescaling to four-dimensional Einstein frame. As a result, in Einstein frame,  there is a direct coupling between $\vo$ and the potentials for both the inflaton $\sigma$ and the quintessence field $\phi$. The total scalar potential describing the dynamics of all three fields is given by $V_{\rm tot} = V_0(\vo) + V_1(\sigma, \vo) + V_2(\phi, \vo)$ where $V_1(\sigma, \vo)$ is generated from the inflaton potential and  $V_2(\phi, \vo)$ from the quintessence potential. Recall that there exists an enormous hierarchy between the energy scales of inflation and dark energy:  $V_{\rm inf} \gtrsim (1\,{\rm MeV})^4 \gg (1\,{\rm meV})^4 \sim V_{\rm DE}$.

During inflation, with $\sigma$ in slow roll, it follows that the quintessence field $\phi$ should be frozen, with $V_1(\sigma,\vo) \gg V_2(\phi, \vo)$. Furthermore, in order to avoid destabilising the volume direction \cite{Kallosh:2004yh}, we need to impose the condition $ |V_0(\vo_*)| \gtrsim V_1(\sigma,\vo) \gg V_2(\phi,\vo)$, where $\vo_*$ is the value of $\vo$ controlling the barrier against decompactification. For Minkowski vacua $\vo_*=\vo_{\rm max}$, the value of $\vo$ at the top of the barrier, while for AdS vacua $\vo_*=\vo_{\rm min}$, the value of the $\vo$ at the minimum\footnote{For dS vacua the story is slightly different: we  need to impose $V_1(\sigma, \vo)\lesssim \left(V_0(\vo_{\rm max})-V_0(\vo_{\rm min})\right)$, although we shall ignore this case since quintessence model building is less well motivated in the presence of a dS vacuum.}. Of course, for AdS vacua inflation is possible only if $V_1(\sigma,\vo)$ acts as an uplifting term such that $V_{\rm inf}  \simeq V_0(\vo_{\rm min}) + V_1(\sigma, \vo_{\rm min})>0$. (Here we are assuming that the location of the minimum, $\vo_\text{min}$, does not change significantly in the presence of the uplift.)

After the end of inflation $V_1(\sigma, \vo)$ goes to zero and so $V_{\rm tot}\simeq V_0(\vo_{\rm min}) + V_2(\phi,\vo_{\rm min})$. For the case where  $V_0(\vo)$ admits a (near) Minkowski vacuum with $V_0(\vo_{\rm min})\simeq 0$,  it follows that $V_{\rm tot}\simeq V_2(\phi, \vo_{\rm min})\simeq V_{DE} \lesssim 10^{-36} V_{\rm inf}\lesssim 10^{-36} V_0(\vo_{\rm max})$, implying a \emph{huge} hierarchy between the energy scales associated with the potential that stabilises $\vo$ and the one which drives quintessence.  For the case where $V_0(\vo)$ admits an AdS vacuum, the hierarchy of scales ensures that, after inflation, $V_{\rm tot}\simeq V_0(\vo_{\rm min}) + V_2(\phi,\vo_{\rm min}) \simeq V_0(\vo_{\rm min}) <0$, implying that quintessence model building is not possible.

Notice that similar considerations would apply if the volume also plays the r\^ole of the inflaton ($\sigma=\vo$), notwithstanding that explicit constructions of volume inflation look rather contrived \cite{Cicoli:2015wja}. Alternatively, if the  volume plays the r\^ole of quintessence ($\phi=\vo$) its potential would, again, be destabilised by the inflationary energy density. Finally if the volume is everything ($\sigma=\phi=\vo$), we would require the presence of two slow roll regions at hierarchically different field values. Given that plateau-like regions can be obtained only by balancing competing terms, if the quintessence epoch at large field values is under control, the inflationary era would lie in a region where perturbation theory would tend to break down. Reheating after the end of inflation and fifth force constraints would also present additional problems in this particular case.

These considerations can be combined with implications of the refined dS conjecture \cite{Ooguri:2018wrx}. The refined dS conjecture rules out quintessence models with a very shallow potential, as in \cite{Cicoli:2012tz}, but allows for quintessence rolling near a hilltop at positive energy (perhaps in the presence of a global AdS or supersymmetric Minkowski vacuum) or down an exponential potential of the form $V=V_0 \ e^{-\lambda \phi/M_p}$. In the latter case, it has been shown that agreement with data requires $\lambda\le 1.02$ at $3 \sigma$ \cite{Akrami:2018ylq}\footnote{A stronger bound of $\lambda \le 0.6$ was obtained in \cite{Agrawal:2018own}. We refer the reader to \cite{Akrami:2018ylq} for a discussion of the two approaches.}. However, our analysis suggests  that these two scenarios are not under better control than dS vacua.

Exponential potentials arise from no-scale breaking perturbative effects for saxions and are typically not small enough to produce the required hierarchy in scales between inflation and dark energy. Therefore, these models are expected to be destabilised by the inflationary dynamics, as well as suffering from problems with the light volume and the F-term. The KL problem also applies to hilltop quintessence near a maximum at positive energy, with a global AdS minimum.
%since this scenario would be plagued by the KL problem. 
We shall present explicit examples of these scenarios and  elucidate their problems in Sec. \ref{sec:KL}. 

To avoid the KL problem, we could consider hilltop quintessence models with a supersymmetric (near) Minkowski vacuum. However, in these models the gravitino mass would be of order the dark energy scale resulting in violation of current bounds \cite{Ferrer:1997yz,Brignole:1997sk, Brignole:1998me,Kawasaki:2008qe}. Moreover, supersymmetric Minkowski solutions are highly  constrained, requiring a very precise form for the superpotential $W$. Therefore, even if the presence of supersymmetry might seem a powerful tool to keep computational control over these solutions, proving their existence in the interior of the moduli space might still be a challenge. As an illustrative example, consider the well-understood type IIB compactifications with $H_3$ and $F_3$ flux, and a  tree-level $W$ that does not depend on the complex volume mode $T=\tau + {\rm i}\theta$. In this set-up, any supersymmetric Minkowski solution at tree-level would necessarily feature a complex flat direction, given by $T$. The existence of a global Minkowski solution with all moduli stabilised would, therefore,  have to rely on the existence of non-perturbative corrections, which lack a full systematic understanding. They would also lift $\tau$ and $\theta$ at the same level of approximation, without generating the right hierarchy between the would-be quintessence field $\theta$ and the volume mode $\tau$.

In the end, we arrive at a generic picture for building a viable quintessence model in string phenomenology. Let us summarize the main points:
\begin{itemize}
\item At leading order (in either perturbative or non-perturbative expansions), the scalar potential $V_0(\vo)$ should feature a (near) Minkowski vacuum with a stabilised volume mode. Notice that non-supersymmetric Minkowski vacua typically require the inclusion of uplifting sectors, and so look qualitatively similar to dS vacua. Although supersymmetric Minkowski solutions could give better computational control, the subdominant effects which generate dark energy would also be responsible for supersymmetry breaking. The gravitino mass (and the soft terms) would not be decoupled from the dark energy scale, in strong tension with both particle physics \cite{Ferrer:1997yz,Brignole:1997sk, Brignole:1998me} and cosmological observations \cite{Kawasaki:2008qe}. Thus the leading order Minkowski vacuum should be \emph{non}-supersymmetric.

\item At the leading order of approximation, the quintessence field should remain flat in order to be able to create the required hierarchy between $V_0(\vo_{\rm max})$ and $V_{\rm DE}$,  with the latter generated by subdominant contributions. The presence of a flat direction can be guaranteed by shift symmetries which fall into two categories: ($i$) non-compact rescaling symmetries for saxions arising from the underlying no-scale structure \cite{Burgess:2014tja, Burgess:2016owb}. However, these are broken by perturbative effects, and so are not  generally efficient enough to provide the required hierarchy; and ($ii$) compact shift symmetries for axions which can potentially generate huge hierarchies, being broken only by tiny non-perturbative effects. Moreover, the smallness of these non-perturbative corrections ensures that the energy density associated with the quintessence potential does not destabilise the volume minimum.

\item Axion quintessence automatically avoids the fifth-force problem (being driven by a pseudo-scalar) and ensures radiative stability thanks to the fact that the axionic shift symmetry is exact at the perturbative level.

\item The main problem with axion quintessence is that its potential is flat enough to drive a period of accelerated expansion only if the axion decay constant is trans-Planckian. However, this situation is very difficult to realise since explicit string constructions with control over the effective field theory tend to have axions with sub-Planckian decay constants \cite{Cicoli:2012sz}, as also implied by the weak gravity conjecture \cite{Arkani-Hamed:2006emk}. There could be counter-examples based on alignment mechanisms \cite{Kim:2004rp, Dimopoulos:2005ac}, although their trustability requires further scrutiny. 

\item For generic axion potentials with sub-Planckian decay constants, we might seek quintessence from a hilltop model. Even if this possibility looks attractive from a model building perspective, we shall see in Sec. \ref{sec:hills} that, when combined with quantum diffusion during the inflationary epoch \cite{Kaloper:2005aj}, it relies on two conditions: ($i$) \emph{very} finely tuned initial conditions; and ($ii$) an extremely low inflationary scale ($H_{\rm inf}\lesssim 1$ MeV), at least for axion decay constants in the regime where the effective field theory is under control.
\end{itemize}
\vskip 0.5cm
\noindent \emph{In other words, from the point of view of theoretical and phenomenological control,  quintessence model building in string theory  is at least as challenging as the search for dS vacua.} 
\vskip 0.5cm
This conclusion raises doubts over the validity of the swampland dS conjecture. Taken alongside the challenges to quintessence, it would imply strong tension between quantum gravity and observation. This might be an indication that phenomenologically relevant solutions to string theory, like dS vacua, lie in the bulk of the moduli space. In this case, it might still be true that perturbation theory is a valid approximation but to be confident of this, we need to refine our technical ability to compute quantum corrections. In the end Nature has already shown an affinity for couplings (as in standard gauge theories and cosmological perturbation theory) that are weak enough to allow us to describe it to a good approximation, even if they cannot be made arbitrarily small.

Finally, if data were to prefer dynamical dark energy, our analysis shows that quintessence models are very unlikely to be axion hilltops since they require highly tuned initial conditions and a very low Hubble scale during inflation. In this regard, axion quintessence models based on alignment mechanisms look more promising even if they need further studies to be convincingly established  in fully fledged string compactifications with moduli stabilisation.

\section{Old challenges for quintessence in string theory} 
\label{sec:modelbuild}

\subsection{Type IIB effective field theory} 

We begin with a brief review of the main techniques for deriving the form of the underlying scalar potential of string compactifications, with a view to building a robust model of dynamical dark energy with all moduli suitably stabilised. More detailed reviews can be found in \cite{Cicoli:2018kdo, AbdusSalam:2020ywo}. We assume that the potential is given by the F-term expression 
\be
V=e^K \left[ K^{i \bar j} D_i WD_{\bar j}\bar W-3 |W|^2 \right],
\ee
where $D_i W=(\partial_i +\partial_i K)W$ is the K\"ahler covariant derivative  and $K^{i \bar j}$ is the inverse of the K\"ahler metric $K_{i \bar j}=\partial_i \partial_{\bar j} K$. Our focus will be on type IIB string compactifications in which the complex structure moduli and the dilaton are fixed at semiclassical level, and so $\partial_i$ denotes partial differentiation with respect to the K\"ahler moduli $T_i=\tau_i+{\rm i} \theta_i$. Even though our focus here is on the effective action of type IIB string theory, our final phenomenological considerations on  quintessence also apply more generally to type IIA and heterotic setups. 

At tree-level, we have a K\"ahler potential, $K=K_0-2 \ln \mathcal{V}$ and a superpotential $W=W_0$, where $\mathcal{V}$ is the volume of the internal Calabi-Yau. $K_0$ and $W_0$ include the complex structure moduli and the dilaton that have already been stabilised, and are therefore assumed to be constant. Because of the `no-scale structure', the corresponding scalar potential vanishes identically. Therefore, to  generate the appropriate masses for the K\"ahler moduli, we must include at least one of the following: ($i$) perturbative corrections to the K\"ahler potential, $K\to K+\delta K_\text{p}$; ($ii$) non-perturbative corrections to the superpotential, $W \to W+\delta W_{\text{np}}$; ($iii$) higher derivative corrections to the scalar potential, $V \to V+\delta V_\text{hd}$.

A general formula giving the K\"ahler moduli dependence of perturbative and higher derivative corrections at all orders in $\alpha'$ and $g_s$ has been provided in \cite{Cicoli:2021rub} exploiting a combination of higher dimensional symmetries such as supersymmetry, scale invariance and shift symmetry, together with techniques from F-theory. This formula reproduces several known explicit computations of quantum corrections. Here we focus on those which have been used for cosmological applications:
\begin{itemize}
\item \textit{$\alpha'^3$ corrections}

These are perturbative corrections in $\ell_s = 2\pi\sqrt{\alpha'}$ to the K\"ahler potential. The leading one arises from $\mathcal{O}(\alpha'^3) \mathcal{R}^4$ terms in the ten-dimensional action and looks like \cite{Becker:2002nn}
\be
K\rightarrow K_0-2\ln\left(\mathcal{V}+{\hat \xi \over 2 }\right)\ ,
\ee
with $\hat \xi=-\frac{\chi(\mathcal{M})\zeta(3)}{2 g_s^{3/2}(2\pi)^3}$ where $\chi(\mathcal{M})$ is the Euler number of the Calabi-Yau $\mathcal{M}$. 

\item \textit{Open string 1-loop corrections}

These are corrections in $g_s$ to the K\"ahler potential, $K\rightarrow K+\delta K_{g_s}$, and are conjectured to take the form \cite{Berg:2007wt, Cicoli:2007xp}
\be
 \delta K_{g_s}=\sum_i g_s {C_i(U,\bar{U}) t_i^{\perp} \over \mathcal{V}}+ \sum_i {\tilde{C}_i(U,\bar{U}) \over t_i^{\cap}\, \mathcal{V}}\ .
\ee
Here there are two contributions: those of $\mathcal{O}(g_s^2 \alpha'^2)$ coming from the tree-level exchange of Kaluza-Klein closed strings, with $t_i^\perp$ denoting the 2-cycles perpendicular to the branes; and those of $\mathcal{O}(g_s^2 \alpha'^4)$ coming from winding strings, with $t_i^\cap$ denoting the 2-cycles of the intersection among branes. $C$ and $\tilde{C}$ are unknown functions of the complex structure moduli $U$, although, as the complex structure sector is fixed at tree-level, one can consider them to be constants. 

\item \textit{Higher derivative corrections}

These are also $\alpha'^3$ corrections to the scalar potential arising from the dimensional reduction of ten-dimensional higher derivative terms of the form $\mathcal{R}^2 G_3^4$, that yield $V \to V+\delta V_\text{hd}$ with \cite{Ciupke:2015msa}
\be
\delta V_\text{hd}=- {g_s^{-3/2}} {3^4 \lambda W_0^4\over \mathcal{V}^4}\,\Pi_i t^i \,,
\label{Vhd}
\ee
where $\lambda$ is an undetermined combinatorial number and $t^i$ are the 2-cycle volume moduli. $\Pi_i$ are topological quantities defined in terms of the $(1,1)$ forms $\hat{D}_i$ as 
\be
\Pi_i=\int_\mathcal{M} c_2\wedge \hat{D}_i\,,
\ee
with $c_2$ the second Chern class of $\mathcal{M}$. Although these effects enter at higher F-term order, they can become important and comparable to string loop corrections.

\item \textit{Non-perturbative corrections}

These are corrections to the superpotential \cite{Blumenhagen:2009qh}
\be
 W \rightarrow W_0+\sum_i A_i\, e^{-a_i T_i}\,,
\ee
related to the existence of E3-brane instantons ($a_i=2\pi$) or gaugino condensates on D7-branes ($a_i=2\pi/N$, where $N$ is the rank of the condensing gauge group). Similar non-perturbative corrections to the K\"ahler potential are subleading when compared to perturbative terms, which can arise from expansions in $\alpha'$ or $g_s$, and will thus be unimportant for our discussion here.
\end{itemize}
Together these corrections yield a scalar potential:
\be
V=\delta V_{\alpha'} +\delta V_{g_s} +\delta V_\text{hd} +\delta V_\text{np}
\ee
where $\delta V_\text{hd}$ is given by \eqref{Vhd} and (setting $e^{K_0}=1$)
\begin{eqnarray}
\delta V_{\alpha'} &=& \frac{3 \hat \xi W_0^2 }{4 \mathcal{V}^3} \label{dValpha} \\
\delta V_{g_s} &=&  \frac{W_0^2}{\mathcal{V}^2} \sum_i\left( g^2_s C_i^2 K^{\text{tree}}_{ii}-2 \frac{\tilde{C}_i}{\vo  t_i^{\cap}}\right) \label{dVgs} \\
\delta V_\text{np} &=& \sum_{i, j} K_{\text{tree}}^{ij}a_ia_jA_iA_j \frac{e^{-a_iT_i-a_j\bar{T}_j}}{\vo^2}+\frac{2 W_0}{\vo^2}\sum_i A_i a_i\tau_i \left(e^{-a_iT_i} +e^{-a_i \bar T_i}\right) 
\end{eqnarray}
$K^{\text{tree}}_{ij}$ is the tree-level contribution to the K\"ahler metric, and $K_{\text{tree}}^{ij}$ its inverse. Notice how the Kaluza-Klein contribution to \eqref{dVgs} enters at second order thanks to the `extended no-scale structure' \cite{Cicoli:2007xp}. All these corrections are under control when the overall volume $\mathcal{V}$ is large. In the regime where all 2-cycles scale as $t \sim \sqrt{\tau}\sim \mathcal{V}^{1/3}$, we have the scaling
\be
\delta V_{\alpha'} \sim {W_0^2\over \mathcal{V}^3}\ , \quad \delta V_{g_s} \sim {W_0^2\over\mathcal{V}^{10/3}}\ ,  \quad \delta V_{\text{np}} \sim {\mathcal{V}^{4/3} e^{-2 a \tau} + W_0 \mathcal{V}^{2/3} e^{-a \tau}\over \mathcal{V}^2}\ , \quad
 \delta V_\text{hd} \sim {W_0^4\over \mathcal{V}^{11/3}}\,, 
\label{eq:scaling_dV} 
\ee
where we have taken $K^{\text{tree}}_{ij} \sim 1/\mathcal{V}^{4/3}$. To generate stable vacua, one has to find a balance between different terms in the potential. For example, in KKLT models \cite{Kachru:2003aw}, $W_0$ is tuned to exponentially small values, $W_0\sim (a\tau) e^{-a \tau}\ll 1$, so that the two contributions to $\delta V_{\text{np}}$ are comparable in size. This typically yields a supersymmetric AdS vacuum whose depth is parametrised by $-W_0^2/\mathcal{V}^2 $. Upon uplift the same scale controls the height of the barrier separating the vacuum from the decompactification limit \cite{Kachru:2003aw}. A notable exception to this rule is the racetrack setup which we describe in Sec. \ref{sec:kl_rev} \cite{Blanco-Pillado:2004aap, Kallosh:2004yh}, where the scale of the vacuum can be made arbitrarily small thanks to two instanton contributions that are aligned relative to one another.  

Another possible approach is to balance perturbative against non-perturbative corrections. Generically we expect the latter to be suppressed, except in the presence of small cycles, as this raises the size of the instanton correction. This is precisely what happens in the LVS scenario \cite{Balasubramanian:2005zx, Conlon:2005ki, Cicoli:2008va}, where $\delta V_{\alpha'}\sim{W_0^2\over \mathcal{V}^3}\sim \delta V_{\text{np}}$. This sets the scale of the potential, controlling both the depth of the non-supersymmetric AdS vacuum and the height of the barrier to infinity which develops after uplifting.   

\subsection{Fifth forces and radiative instability}

Some dynamical dark energy models have already been built within the framework of string compactifications \cite{Choi:1999xn, Kaloper:2008qs, Panda:2010uq, Cicoli:2012tz, Kamionkowski:2014zda, Olguin-Trejo:2018zun}. Typically the quintessence field corresponds to the lightest mode and the other moduli are stabilised at tree-level and by leading order corrections. In this way dynamical dark energy appears as a next-to-leading order effect, allowing us to retain perturbative control. It also guarantees that the slow roll of the quintessence field away from the minimum only displaces the volume mode from its original vacuum expectation value by a small amount. 

However none of the existing quintessence models in the literature is really satisfactory due to several challenges which were already highlighted in \cite{Cicoli:2018kdo, Hebecker:2019csg}. These challenges are related to the phenomenological requirements that a prospective stringy quintessence field would have to satisfy, namely:
\begin{enumerate}
\item A \textit{light quintessence modulus} $\phi$ with $m_\phi\lesssim H_0 \sim 10^{-60}\,M_p$. This follow directly from requiring that the scalar field $\phi$ is in slow roll at the current epoch.

\item \textit{Heavy superpartners} with masses $M_{\rm soft}\gtrsim10^{-15}\,M_p$. Supersymmetric partners must be above the threshold set by the LHC \cite{Zyla:2020zbs}. This, in turn, yields large perturbative corrections from loops of visible sector supersymmetric particles.

\item \textit{Heavy Kaluza-Klein scale} with $M_{\rm KK}\gtrsim10^{-30}\, M_p$. Sub-millimetre scale tests of Newtonian gravity put a bound on the Kaluza-Klein scale \cite{Kapner:2006si}. 

\item \textit{Heavy volume modulus} with $m_{\mathcal{V}}\gtrsim 10^{-30}\,M_p$. Upon compactification, the four-dimensional Ricci scalar gets a prefactor which depends on the volume modulus which couples to matter fields after Weyl rescaling to Einstein frame. There are stringent bounds on such fifth force effects given by sub-millimetre experiments \cite{Kapner:2006si, Damour:2010rp, Acharya:2018deu}.
\end{enumerate}

The authors of \cite{Hebecker:2019csg} discuss the implications of these requirements for string models of dark energy, with a focus on LVS-motivated scenarios for concreteness. Two main issues arise.

\paragraph{The light volume problem:} 

The Kaluza-Klein mass is given by
\begin{equation}
    M_{\rm KK}={M_s\over R}\sim \frac{M_p}{\mathcal{V}^{2/3}}\gtrsim 10^{-30}\,M_p\qquad \Rightarrow \qquad \mathcal{V}\lesssim 10^{45}\ , \label{eq:mkk_bound}
\end{equation}
where we have used $M_s\simeq M_p\,\mathcal{V}^{-1/2}$, the fact that the radius of the compact space $R\sim \mathcal{V}^{1/6}$, and the bound on the Kaluza-Klein mass given above.

In the LVS scenario, the mass for the volume modulus is generated through leading $\alpha'$ corrections \eqref{dValpha}, while at subleading order loop corrections \eqref{dVgs} lift additional K\"ahler moduli which could play the r\^ole of the quintessence field $\phi$. Using \eqref{eq:scaling_dV} and \eqref{eq:mkk_bound}, one finds
\begin{equation}
    \frac{m_\phi}{m_\mathcal{V}}\sim \sqrt{\frac{\delta V_{g_s}}{\delta V_{\alpha'}}}\sim \frac{1}{\mathcal{V}^{1/6}} \gtrsim 10^{-7}\ .
    \label{eq:modVol}
\end{equation}
In \cite{Cicoli:2012tz} loop contributions are suppressed due to low energy supersymmetry in the bulk and an anisotropic shape of the extra dimensions. The quintessence field $\phi$ is instead lifted by poly-instanton effects which give
\begin{equation}
    \frac{m_\phi}{m_\mathcal{V}}\sim \sqrt{\frac{\delta V_{\rm poly}}{\delta V_{\alpha'}}}\sim \frac{1}{\sqrt{\vo}} \gtrsim 10^{-22}\ .
    \label{eq:modVol2}
\end{equation}
However, both (\ref{eq:modVol}) and (\ref{eq:modVol2}) are in contradiction with the phenomenological bound imposed by fifth force constraints and the value of $H_0$, i.e. ${m_\phi \over m_{\mathcal{V}}}\lesssim 10^{-30}$. A way to avoid this issue is to introduce subleading effects that modify the volume scaling of \eqref{eq:modVol} and (\ref{eq:modVol2}) \cite{Cicoli:2011yy}. For example, in the model of \cite{Cicoli:2012tz}, $\phi$ does not mediate any fifth force since its coupling to Standard Model fields is weaker than Planckian due to sequestering effects in the extra dimension (see also \cite{Acharya:2018deu} for estimates of moduli couplings in sequestered models with large extra dimensions). Nevertheless, the volume mode would lead to new long range interactions since, due to (\ref{eq:modVol2}), it is much lighter than 1 meV and it couples with ordinary matter with standard Planckian strength (however, see \cite{Burgess:2021qti} for a possible screening effect due to the kinetic coupling of $\vo$ to its associated axionic field). 

\paragraph{The F-term problem:}  

The mass of the superpartners, which we approximate by the gaugino mass, is of the order
\begin{equation}
    M_{\rm soft}\sim M_{1/2} = \frac{F^i \partial_i f}{{\rm Re}(f)}\,M_p\,,
\end{equation}
where $f$ is the gauge kinetic function. If we assume that supersymmetry breaking is mediated through some higher-dimensional operator at some scale $M_b$, for a simple toy model with a single spurion field $X$ and F-term $F_X$, the contribution to the scalar potential is 
\begin{equation}
    \delta V_X \sim F_X^2 \sim M_b^2 M_{\rm soft}^2\ ,
\end{equation}
where $M_{\rm soft}$ enters the scalar potential after canonical normalisation of the spurion term. If we require that supersymmetry breaking is mediated above the TeV scale, together with the phenomenological constraints on the superpartner masses, we find $\delta V_X\gtrsim 10^{-60} M_p\gg H_0^2$. This contribution would raise the scale of the potential well beyond the  dark energy scale. A loophole is to consider a new contribution to the scalar potential that would cancel supersymmetry breaking effects with some fine-tuning, as in \cite{Cicoli:2012tz}, where the additional effect is assumed to come from the backreaction of  non-supersymmetric visible sector branes (see also \cite{Burgess:2021obw} for recent developments of quintessence models in scenarios with non-linearly realised supersymmetry).

The challenges for quintessence outlined in \cite{Cicoli:2018kdo, Hebecker:2019csg} are just the tip of the iceberg. In the next section, we identify an even bigger problem: disruption of the energetic dynamics by the inflationary energy density, resulting in destabilisation of the volume mode and decompactification. 

\section{The KL problem for quintessence}
\label{sec:KL}

In \cite{Kallosh:2004yh} Kallosh and Linde argued that the scale of inflation is bounded from above by the gravitino mass in the standard KKLT scenario \cite{Kachru:2003aw}.  The constraint arises in order to avoid a runaway in the volume mode, leading to  decompactification at early times. Similar considerations were used to place limits on thermal corrections to the scalar potential, imposing a maximum temperature in the four-dimensional effective theory \cite{Buchmuller:2004tz, Anguelova:2009ht}. The KL problem extends beyond KKLT, and has also been shown to affect LVS models where the constraint turns out to be even stronger \cite{Conlon:2008cj}. We begin by reviewing the key aspects of the original argument of \cite{Kallosh:2004yh}. Later we will show that it has implications also for string models of dynamical dark energy. 

\subsection{Review of the KL problem}
\label{sec:kl_rev}

Consider a one-instanton KKLT model with superpotential and $\alpha'$-corrected K\"ahler potential given by
\be
W=W_0+A\,e^{-a T}\,, \qquad K=K_0 -2\ln\left(\mathcal{V}+\frac{\hat{\xi}}{2}\right),
\ee
where $\mathcal{V}=(T+\bar T)^{3/2}$ is the volume of the internal Calabi-Yau manifold with $T=\tau+{\rm i} \theta$ the 4-cycle volume modulus. The dynamics of the moduli, to leading order in $\hat{\xi}$, is given by the Lagrangian
\begin{equation}
    \mathcal{L}=K_{T\bar{T}} \,\partial T \partial \bar{T} - V(\tau,\theta) = \frac{3}{4\tau^{2}}\left(1-\frac{5\hat{\xi}}{2^{7/2}\tau^{3/2}}\right)\left[(\partial\tau)^2+(\partial\theta)^2\right]-V(\tau,\theta)\ ,
\end{equation}
where the F-term potential is
\be \label{Vkklt}
V_\text{KKLT}(\tau, \theta)=\frac{a^2A^2e^{-2a\tau}}{6\tau}\left(1+\frac{3}{a\tau}\right)-\frac{aA|W_0|e^{-a\tau}}{2\tau^2}\cos (a \theta)+\frac{3 W_0^2\hat{\xi}}{64 \sqrt{2}\tau^{9/2}}\,,
\ee
and, without loss of generality, we have assumed $W_0$ to be real and negative, $W_0 = -|W_0|$. It then follows that a supersymmetric minimum exists at zeroth order in $\xi$, located at $\theta=0$ and  $\tau=\tau_{\text{min}}$,  where the latter satisfies the following relation
\be \label{W0kklt}
|W_0|=A\,e^{-a\tau_{\text{min}}}\left( 1 +\frac{2}{3}a\tau_{\text{min}}\right)\simeq \frac23 A a \tau_{\rm min} \,e^{-a\tau_{\text{min}}}\,.
\ee
In this section, we follow \cite{Kallosh:2004yh} and concentrate on the dynamics close to the minimum, neglecting the subleading effects of $\hat\xi$. Of course, at large volumes, these $\alpha'^3$ corrections will induce a maximum in the potential as stressed in \cite{Conlon:2018eyr}, which will be relevant to the discussion in Sec. \ref{DetailedImplications}. In the past these very same corrections have been used for uplifting the supersymmetric AdS minimum to Minkowski \cite{Westphal:2006tn}, a more extreme regime which we will not consider here. 

\begin{figure}[H] 
	\centering
	\begin{minipage}{0.45\textwidth}
		\centering
		\includegraphics[width=75 mm]{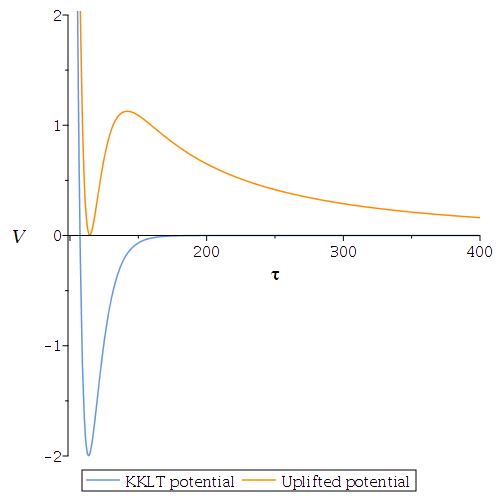}
	\end{minipage}\hfill
	\begin{minipage}{0.45\textwidth}
		\centering
			\includegraphics[width=75 mm]{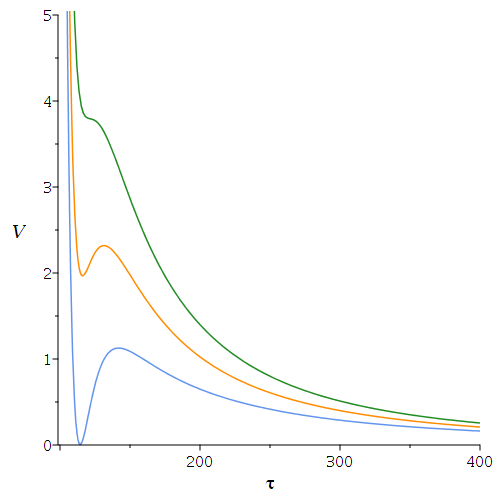}
	\end{minipage}\hfill
\caption{Left: KKLT scalar potential with and without uplift. Right: Uplifted potential with increasing inflationary corrections.}	\label{kklt.fig}
\end{figure}

It follows that the potential at the minimum is AdS, and given by 
\be \label{Vads}
V_{\rm AdS} \equiv V_\text{KKLT}(\tau_{\text{min}}, 0)=-\frac{a^2A^2e^{-2a\tau_{\text{min}}}}{6\tau_{\text{min}}}=- 3\left(\frac{ |W_0|}{\vo_{\rm min}}\right)^2.
\ee
The AdS vacuum can be uplifted to a metastable dS vacuum, for example, using  a warped anti-D3 brane, which gives a contribution to the scalar potential of the form
\be
V_{\text{up}} = \frac{C}{\tau^2} \,.
\ee
When the axion is at its minimum, the corresponding KKLT potential for the saxion is given by $V_{\text{KKLT}'}(\tau) =V_\text{KKLT}(\tau, 0) +V_{\text{up}}(\tau)$. The uplift is tuned so that the new dS minimum, located at $\tau_\text{dS}$, is compatible with current bounds on the cosmological constant, that is $V_{\text{KKLT}'}(\tau_\text{dS}) =V_\text{KKLT}(\tau_\text{dS}, 0) +V_{\text{up}}(\tau_\text{dS}) \lesssim 10^{-120}\,M_p^4$. For the instanton expansion to be under control at the minimum, it must be placed at some large value of $\tau$. As a consequence, the uplift does not have a huge effect on its position, and we can take  $\tau_{\text{dS}}\simeq\tau_{\text{min}}$. It follows that the scale of the uplift is simply given by the scale of the AdS vacuum, as one might already have expected,  $V_{\text{up}}(\tau_\text{dS}) \approx |V_\text{AdS}|$. Furthermore, since the original scalar potential decays exponentially quickly in comparison to the uplift term at large $\tau$, the metastable vacuum is separated from the runaway region by a barrier whose height is fixed  at the same scale, $V_*\sim V_{\text{up}}(\tau_\text{dS})\approx \left|V_{\text{AdS}}\right|$. The generic shape of the potential with and without uplift is shown in \figref{kklt.fig}.

With supersymmetry now broken by the anti-D3 brane, the gravitino acquires a mass at the uplifted minimum, given by 
\be
m^2_{3/2}=\left[e^K|W|^2\right]_{\tau=\tau_{\text{dS}},\theta=0}\approx \left[e^K|W|^2\right]_{\tau=\tau_{\text{min}},\theta=0} = \left(\frac{ |W_0|}{\vo_{\rm min}}\right)^2 = \frac{|V_\text{AdS}|}{3}
\ee
where we have used the fact that $D_T W=0$ and so $V_\text{AdS}=-3e^K|W|^2$ for the supersymmetric AdS vacuum. 

The KL problem emerges when we consider inflation in this particular setup. To begin with, one could consider hilltop inflation from the top of the potential barrier $V_*$. In this case, the Hubble parameter during inflation is related to the gravitino mass at present through
\be
H_{\rm inf}^2 \approx  \frac{V_*}{3} \sim \frac{\left|V_{\text{AdS}}\right|}{3} \approx  m_{3/2}^2\ . \label{eq:H1}
\ee
Another mechanism for inflation could be due to the dynamics of branes in the compact space \cite{Hsu:2003cy}. In this case, the inflaton is some other modulus field $\sigma$ controlling the location of the D-branes in the internal space. The uplifted KKLT potential then receives a contribution from the inflaton due to the structure of the supergravity F-term potential. The inflationary potential will generically take the form
\be
V_{\text{inf}}(\tau)= V_{\text{KKLT}'}(\tau)+\frac{V(\sigma)}{\tau^3}\ . \label{eq:KL_pot}
\ee
As shown in \figref{kklt.fig}, the inflationary corrections raise the level of the minimum more than they raise the height of the barrier. Eventually, if the scale of inflation is high enough, the local minimum turns into an inflection point and the barrier disappears completely. 
For the volume modulus to stay stabilised during inflation, we require that $V_{\text{inf}}\lesssim V_* \sim 3 m_{3/2}^2 M_p^2$, and again we find a relation between the value of the Hubble parameter during inflation and the gravitino mass today
\be
H_{\rm inf}^2 \approx  \frac{V_\text{inf}}{3 M_p^2} \lesssim m_{3/2}^2\ . \label{eq:H2}
\ee
Equations \eqref{eq:H1} and \eqref{eq:H2} are the main results of \cite{Kallosh:2004yh} and show that, unless one can parametrically decouple the value of the gravitino mass from the height of the potential barrier, the scale of inflation is bounded from above by the gravitino mass, $H_{\rm inf} \lesssim m_{3/2}$. This presents a problem in that it sets the gravitino mass to be extremely large, which by proxy sets the scale of supersymmetry breaking to be much larger than the TeV scale. Notice that this is not necessarily a problem for models with a high scale of supersymmetry breaking. However, to obtain the observed value of the Higgs mass, these require severe fine tuning  or a sequestered visible sector on D3 branes at singularities where $M_{\rm soft}\ll m_{3/2}$ \cite{Aparicio:2014wxa}. The tension between low scale supersymmetry and inflation is even more acute in LVS models since the barrier is generically not as high, scaling as $V_* \sim m_{3/2}^3 M_p$. Using similar arguments as above, this constrains the scale of inflation to be $H_{\rm inf} \lesssim m_{3/2} \sqrt{m_{3/2}/M_p}$.

\subsection{Racetrack solution to the KL problem}

\begin{figure}[H]
		\centering
		\includegraphics[width=75 mm]{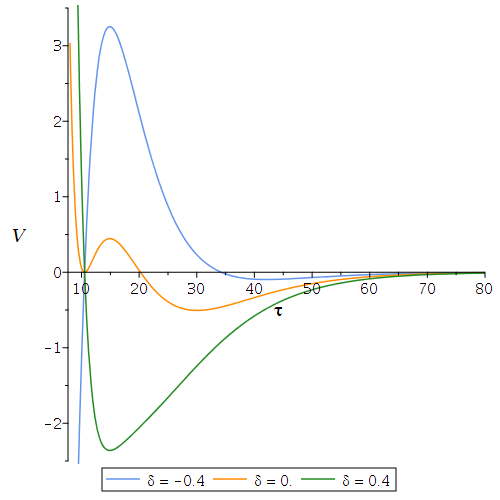}
	\caption{Racetrack potential at $\theta=0$ for parameter choice $
    A = 1, B = -1, a = 0.1, b = 0.09$
and different values of $W_0=W_0^{\text{crit}}(1+\delta$).} \label{racetrack.fig}
\end{figure}

Kallosh and Linde \cite{Kallosh:2004yh} proposed a resolution to this problem within  a racetrack model \cite{Blanco-Pillado:2004aap}, where the superpotential receives a second instanton contribution
\be
W=W_0+A\,e^{-a T}+B\,e^{-b T}\,,
\ee
as one would expect from gaugino condensation in a theory with a product gauge group. In particular, for $SU(M) \times SU(N)$ we expect $a=2\pi/M$ and $b=2\pi/N$. The corresponding F-term potential is given by
\begin{eqnarray}
V_\text{race}(\tau, \theta)&=&\frac{a^2A^2e^{-2a\tau}}{6\tau}\left(1+\frac{3}{a\tau}\right)+\frac{b^2B^2e^{-2b\tau}}{6\tau}\left(1+\frac{3}{b\tau}\right)  \nonumber \\
&+&\frac{abAB e^{-(a+b)\tau}}{3\tau} \left(1+\frac{3}{2a\tau}+\frac{3}{2b\tau}\right)\cos[(a-b)\theta]
 \nonumber\\
&-& \frac{aA|W_0|e^{-a\tau}}{2\tau^2}\cos (a \theta)-\frac{bB|W_0|e^{-b\tau}}{2\tau^2}\cos (b \theta) \,. 
\label{raceV}
\end{eqnarray}
The model admits a supersymmetric Minkowski vacuum for a critical value of $W_0$, given by 
\be
|W_0|^{\text{crit}}=A\,\mathcal{R}^\frac{a}{b-a}+B\,\mathcal{R}^\frac{b}{b-a}
\ee
where we assume $\mathcal{R}=-\frac{a A}{b B}>1$ and $a>b$ for definiteness. The minimum is located at $\theta=0$ and  $\tau_{\rm min}=\frac{1}{(a-b)} \ln \mathcal{R}$. The gravitino mass vanishes at the Minkowski vacuum since supersymmetry remains unbroken. As such  there is no relation between the gravitino mass and the height of the potential barrier. The latter scales as $V_* \lesssim (a-b)^3 A^2 \mathcal{R}^\frac{2a}{b-a}$, provided we assume $(a-b) \ll 1$, so that the minimum is pushed to a large value of $\tau$. The shape of the potential for different values of $W_0$ is shown in Fig. \ref{racetrack.fig}.

\subsection{General implications for quintessence}
\label{implications}

Although \cite{Kallosh:2004yh} were interested in constraining the scale of inflation and the form of the inflationary potential,  similar considerations can be applied to the low energy potential describing the dynamics of dark energy today.  We focus on the dynamics of up to three moduli: a quintessence field $\phi$ describing dark energy, an inflaton $\sigma$ and the volume modulus $\tau$. In principle, these could be three different moduli, or they could overlap - we consider all possibilities. 

We start by assuming they are all different. In general, the full scalar potential can be written as
\be
V_{\rm tot}(\phi, \sigma, \tau)=V_0(\tau)+V_1(\sigma, \tau)+V_2(\phi, \tau)
\ee
where $V_0$ is the potential that fixes the volume mode, $V_1$ the contribution of the inflaton and $V_2$ from quintessence.
In this case, dark energy is assumed to be described by a scalar potential $V_{\rm DE}(\phi, \tau)=V_0(\tau)+V_2(\phi, \tau)$ with $\phi$ in slow roll at some scale $\phi \sim \phi_0$ today and the volume stabilised at some large value $\tau =\tau_0$. 
However, during inflation, we generically expect the full scalar potential to receive an inflaton-dependent correction as described in the original KL scenario \cite{Kallosh:2004yh}. In other words, 
\be
V_{\rm inf}(\phi, \sigma, \tau)=V_{\rm DE}(\phi,\tau)+\frac{V(\sigma)}{\tau^3}\ .
\ee
where we have set $V_1(\sigma, \tau)=\frac{V(\sigma)}{\tau^3}$, as a result of Weyl rescaling. The inflaton field rolls slowly through at least $50$ efoldings of inflation, starting out at $\sigma_\text{inf}$ and ending at $\sigma_0$, with $V(\sigma_0) \approx 0$. The volume modulus and the quintessence field are assumed to be spectators during inflation, stabilised at $\tau \approx \tau_\text{inf}$ and $\phi \approx \phi_\text{inf}$ respectively. Once inflation has ended, the inflaton dumps energy into the Standard Model sector. During this phase of reheating, the volume and quintessence moduli are allowed to move, if necessary, towards their current values, $\tau \to \tau_0$, $\phi \to \phi_0$. However, in order to avoid potential problems with light element abundances \cite{Pitrou:2018cgg} and the spectrum of the cosmic microwave background radiation \cite{Planck:2018vyg}, all three moduli must remain stabilised from nucleosynthesis onwards, right up until the current epoch of dark energy domination, at which point $\phi$ starts to slow roll.  

The key observation is that 
$V_{\rm DE}(\phi_0,\tau_0)\simeq H_0^2 \ll H_{\text{inf}}^2 \simeq V(\sigma_{\text{inf}})/\tau_{\text{inf}}^3$, 
where $H_0$ is the current Hubble scale, and $H_{\rm inf}$ is the scale of inflation. The hierarchy is a considerable one: the scale of dark energy is $H_0 \sim 10^{-60} M_p$, whereas the scale of inflation is assumed to lie somewhere in the range $10^{-42} M_p \lesssim H_\text{inf} \lesssim 10^{-5} M_p$.\footnote{Here, the lower bound comes from the scale of BBN (around MeV), although the actual temperature of the primordial bath might be higher, of $\mathcal{O}(\text{GeV})$ \cite{Davidson:2000dw}. The upper bound comes from constraints on the tensor-to-scalar ratio \cite{Planck:2018vyg}.} The considerations of \cite{Kallosh:2004yh}, now suggest a parametric separation in the scale of the underlying quintessence potential in the early universe and at late times. In particular, we can constrain the scale at early times, given by $V_{\rm DE}(\phi_\text{inf},\tau_\text{inf})$ by demanding that the volume modulus remains stabilised during inflation.  Following the same logic as \cite{Kallosh:2004yh}, we note that to avoid the runaway in the volume, the corresponding minimum at $\tau_\text{inf}$ should be separated from the asymptotic region by a barrier, $V_*$, as high as the scale of inflation, $V_* \gtrsim H^2_\text{inf}$. Given that generically we expect $V_* \sim |V_{\rm DE}(\phi_\text{inf},\tau_\text{inf})|$, to avoid a runaway we require $|V_{\rm DE}(\phi_\text{inf},\tau_\text{inf})| \sim V_* \gtrsim  H^2_\text{inf} \gg H_0^2 \simeq V_{\rm DE}(\phi_0,\tau_0)$.  As we will see in a moment, it is hard to see how we can achieve this separation of scales in a controlled setup.

One of the lessons from Sec. \ref{sec:modelbuild} is that, at leading order and weak coupling, one scale typically controls the scale of the AdS vacuum and the height of the barrier, both going as $W_0^p/\mathcal{V}^q \ll 1$, for some $p, ~q$. If the volume, inflaton and quintessence fields, correspond to three different moduli, we have seen how the considerations of Kallosh and Linde \cite{Kallosh:2004yh} suggest that the barrier height should be at least as large as the scale of inflation to avoid a runaway. This fixes the scale of the underlying potential to be far in excess of the dark energy scale, $W_0^p/\mathcal{V}^q \gtrsim H_\text{inf}^2 \gg H_0^2$. The AdS vacuum, even if it could be uplifted to Minkowksi by the inflationary energy density, would now be too deep for any next to leading order correction to be a viable dark energy candidate, where the potential must be positive.

Although our arguments have focused on the case where the inflation, the volume and quintessence field are three different moduli, the situation is not improved when we relax this assumption. Let us consider each of the alternatives:
\begin{itemize}
\item \textit{The volume accounts for dark energy but not for inflation:} The inflaton is once again assumed to be some other (s)axion orthogonal to the volume mode. It is then required that the volume mode stays stabilised during inflation and finds itself in a gentle slope at late times, giving rise to dynamical dark energy. However, in order to avoid the runaway during inflation, the potential must have a large barrier, far in excess of the scale of the late time potential. This scenario is very similar to the one we have already described, and as such, suffers from the same difficulties. Volume driven quintessence  will also give rise to long range forces that violate fifth force constraints (see e.g. \cite{Will:2014kxa}).

\item \textit{The volume accounts for inflation but not for dark energy:} In this scenario, the potential for the volume contains a high scale plateau, allowing the volume to roll slowly during inflation. After inflation, the volume should settle into a low scale Minkowksi vacuum. This could then be stabilised at leading order, with some next-to-leading effect giving rise to dynamical dark energy through another modulus. In \cite{Conlon:2008cj,Cicoli:2015wja} volume inflation near an inflection point has been realised by considering different competing contributions: non-perturbative effects, string loops, higher derivative corrections, anti-branes and charged hidden matter fields. Besides looking very contrived and tuned, these constructions raise doubts as to the level of perturbative control since the value of the volume during inflation is relatively small. Moreover, one should make sure that the quintessence field away from the minimum does not result in the volume being destabilised.\footnote{Notice that the tension between $H_{\text{inf}}$ and $H_0$ could be relaxed by  also having $W_0$ evolve from large to small values during inflation, as in the toy model of \cite{He:2010uk}. However we are not aware of a robust model that realises this effect while remaining under computational control.}
 
\item \textit{The volume accounts for everything:} In this case, the volume modulus is responsible for both inflation and dark energy. The situation is similar to the previous case but now we require a flat enough plateau later on as well. This seems to require competing terms at both small volumes (during inflation) and large volumes (during the dark energy period), with a significant hierarchy built in. Such hierarchies would need to be generated by exponentials, which are generated non-perturbatively. This suggests the early time behaviour may not be under perturbative control. Furthermore, if inflation ends with the inflaton rolling in a steep potential, and not approaching a minimum, reheating would need to be non-standard. Crucially the late time behaviour would also fall foul of fifth force constraints.
\end{itemize}

We can try to get around these  problems by assuming that the stabilisation of the volume lies at some low scale, near Minkowski  vacuum generated at leading order, breaking the connection between the scale of the vacuum and the height of the barrier. (Recall that the barrier height should exceed the scale of inflation to avoid decompactification.) If this leading order stabilisation leaves, say,  an axionic flat direction which is lifted only at subdominant order by tiny non-perturbative effects, one could reproduce the required hierarchy between $H_{\rm inf}$ and $H_0$ without inducing any destabilisation of the volume mode. Notice, however, that generating a supersymmetric Minkowski minimum ($W=0$) by solving the F-terms equations ($D_i W=0$) requires a finely tuned cancellation between all contributions to the superpotential, both at tree and non-perturbative level, as in the  racetrack scenario \cite{Blanco-Pillado:2004aap}, which was already identified as a way to skirt around the original KL problem \cite{Kallosh:2004yh}. Moreover, axion quintessence in agreement with swampland bounds on the associated decay constant, requires dynamical dark energy to occur close to the maximum of the axion potential where the scale of supersymmetry breaking would be extremely low, set by the  scale of dark energy. This leads us to conclude that {\it non}-supersymmetric Minkowski vacua are actually more appealing, at least if we want to build a viable  model of quintessence in string theory.

\subsection{A closer look at quintessence models with a KL problem}
\label{DetailedImplications}

When we consider quintessence in string theory, commitment to the refined dS and weak gravity conjectures forbids a dynamical model of dark energy based on either of the following scenarios: 
\begin{enumerate}
\item Minkowski vacuum with saxion slow roll down a very shallow potential \cite{Cicoli:2012tz}
\item Minkowski vacuum with axion quintessence with trans-Planckian decay constant \cite{Arkani-Hamed:2006emk}
\end{enumerate}
Whilst this leaves some alternatives, our consideration of the KL problem in the previous section suggests that most of these are also ruled out. In particular, the following scenarios 
\begin{enumerate}
\item Saxion hilltop for a Minkowski or AdS vacuum
\item Axion hilltop for a Minkowski vacuum with no hierarchy
\item Saxion slow roll down a moderate slope, with a runaway or a Minkowski vacuum
\end{enumerate}
are all compatible with the dS conjecture. Two minor clarifications are in order here. By `hierarchy' we mean the existence of an exponential hierarchy of scales between the leading order potential for the volume and the axion potential responsible for quintessence. By a `moderate slope' we mean order one in Planck units, {\it i.e.} steep enough to satisfy the refined dS conjecture but shallow enough to allow for at least one efolding of slow roll. 

Each of these three alternatives suffers from the KL problem. They also suffer from a variety of other problems, not least that of an unacceptably light volume modulus and a light gravitino. In this section we study specific examples of each scenario, explicitly demonstrating how many of these problems emerge.

\subsubsection*{Saxion hilltop for a Minkowski or AdS vacuum}

The racetrack scenario has a  supersymmetric Minkowski vacuum, separated from the runaway regime by a maximum in the volume mode. We can therefore imagine a dynamical model of dark energy where  the volume mode is rolling close to the hilltop, and the axion is fixed at its minimum, at $\theta=0$.  As we saw previously, the racetrack scenario was proposed as a way around the original KL problem, since the height of the barrier can be taken to be higher than the scale of inflation without any consequence on the gravitino mass. However, the height of the barrier is the height of the maximum in the volume direction (or better, the height of saddle in the ($\tau, \theta$)-plane). For hilltop quintessence driven by the volume mode, this height is now set by the dark energy scale
\begin{equation}
H_0^2 \sim V_\text{race}(\tau_{\rm max}, 0)\,.   
\end{equation}
Clearly this barrier is too small to protects us from the KL problem. Indeed, the contribution from inflation,  driven by a different field $\sigma$,  couples to the volume mode due to Weyl rescaling.  As expected, it will induce  destabilisation of the volume towards decompactification since 
\begin{equation}
\frac{V(\sigma)}{\tau^3} \sim H_{\rm inf}^2  \gg H_0^2 \sim V_\text{race}(\tau_{\rm max}, 0)  
\end{equation}

Of course, similar considerations also apply  to saxion hilltops where the global minimum is supersymmetric AdS. As an example, consider  KKLT models where a hilltop in the volume modulus is generically present, even in the absence of an anti-D3 brane uplift, as a consequence of $\alpha'^3$ corrections to the K\"ahler potential, as already stressed in \cite{Conlon:2018eyr}.  This model is also tractable  enough to easily demonstrate other issues that can emerge beyond the KL problem, such as the light volume modulus and the light gravitino. Let us run through some of the details.

The KKLT potential was already given in \eqref{Vkklt}. If we assume that the axion $\theta$ is stabilised at its minimum at $\theta=0$, the dynamics of the volume modulus $\tau$ in a neighbourhood of the maximum is controlled by the last two terms, in other words
\be
V_\text{KKLT}(\tau, 0) \approx -\frac{aA|W_0|\,e^{-a\tau}}{2\tau^2}+\frac{3 W_0^2\hat{\xi}}{64 \sqrt{2}\tau^{9/2}}\ ,
\ee
with $|W_0|$ given by \eqref{W0kklt}. This simplification allows us to show that the maximum is located at $\tau_\text{max}$, defined  by the relation
\be
\hat \xi=\frac{64 \sqrt{2} a A \tau_\text{max}^{5/2} e^{-a \tau_\text{max}} (a \tau_\text{max}+2)}{27 |W_0|}.
\ee
Since $a\tau_\text{max} \gg 1$, the height of the potential at the maximum is 
\be \label{Vkkltmax}
V_\text{KKLT}(\tau_\text{max}, 0)\approx \frac{ a^2 A |W_0|\,e^{-a \tau_\text{max}}}{9 \tau_\text{max}} = \frac{4}{9} a \tau_\text{min}\left(\frac{\tau_\text{min}}{\tau_\text{max}}\right) e^{-a \Delta \tau}|V_\text{AdS}|
\ee
where we made use of \eqref{W0kklt} and the expression for the scale of the leading order AdS minimum \eqref{Vads}. The minimum at $\tau_\text{min}$ and the maximum at $\tau_\text{max}$ are separated by distance $\Delta \tau= \tau_\text{max}-\tau_\text{min}>0$. 

Clearly the height of the maximium should be  fixed by the current Hubble scale, $H_0^2 \sim V_\text{KKLT}(\tau_{\rm max}, 0) $. By the same reasoning as for the racetrack scenario, we run into a KL problem.  In this simple model, it is also instructive to demonstrate the smallness of the mass of the gravitino and the volume modulus explicitly.

Current observational bounds require $V_\text{KKLT}(\tau_\text{max}, 0)\sim 10^{-120}$ in Planck units. Such low values can be achieved either by having $\tau_{\rm max}$ large or 
$|W_0|$ exponentially small, two requirements that are not independent in KKLT, as can be seen from \eqref{W0kklt}. 
In order to estimate the choice of  parameters that leads to the correct value for $H_0$, we use the fact that the maximum and the minimum are not too far apart, and  compute the height of the maximum to zeroth order in $\Delta \tau$
\be
V_\text{KKLT}(\tau_\text{max}, 0)\approx  \frac{4}{9} a \tau_\text{min}|V_\text{AdS}|\approx\frac{2 a^3 A^2 }{27}e^{-2 a\tau_\text{min}}
\ee
where we made use of \eqref{Vads}. Assuming $\frac{2 a^3 A^2 }{27}=\mathcal{O}(1)$, matching the observed value of $H_0$ requires $a \tau_\text{min}\sim 140$, which through \eqref{W0kklt} translates as $|W_0|\sim 10^{-59}$ and a gravitino mass of $m_{3/2}\sim 10^{-33}$ eV. This is unacceptably light \cite{Ferrer:1997yz,Kawasaki:2008qe}. The fact that the gravitino mass is of order the dark energy scale can be traced back to the fact that the leading order vacuum is supersymmetric and very close to Minkowski. Indeed, from equation \eqref{Vkkltmax}, we see that the scale of the supersymmetric AdS vacuum is bounded above by the dark energy scale. This failure to decouple $m_{3/2}$ and $H_0$ is clearly typical of any model featuring a leading order  supersymmetric Minkowski, or near Minkowski,  vacuum. 

To compute the mass of the volume mode, one has first to switch to a canonical field via $\phi = \sqrt{\frac32} \ln \tau$ and then compute $m_\phi^2\simeq V_{\phi\phi}$ at the location of the maximum. This yields $m_\phi^2 \simeq -3 a \tau_\tmax V_0 \simeq -3 a \tau_\tmax m_{3/2}^2$. Since  $a\tau_\text{max} \gtrsim a\tau_\text{min}\sim 140$, this implies that the mass of the volume mode is only one order of magnitude above the gravitino mass, explicitly showing the existence of a light volume problem.

\subsubsection*{Axion hilltop for a Minkowski vacuum with no hierarchy}

Let us return to the racetrack model and consider using $\theta$, instead of $\tau$, to drive quintessence. Once again, since the dark energy scale now sets the scale of the potential, this will immediately  run into a KL problem.  As it happens, this model suffers from another problem, closely related to the KL problem, but applied only to late time dynamics. Indeed, even if we ignore the contributions from inflation, the volume barrier disappears as soon as we move the axion sufficiently far away from its minimum. In other words, in attempting to move the axion to the hilltop, the volume itself is immediately destabilised. 

\begin{figure}[H]
		\centering
		\includegraphics[width=75 mm]{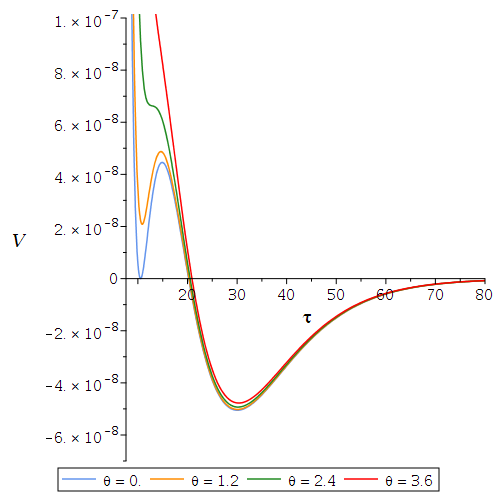}
	\caption{Racetrack potential at different values of the axion $\theta$ for parameter choice $
    A = 1, B = -1, a = 0.1, b = 0.09$
.} \label{racetrackbarrier.fig}
\end{figure}

This is demonstrated numerically in \figref{racetrackbarrier.fig}. Here we plot the form of the racetrack potential \eqref{raceV} as a function of the volume modulus $\tau$ for different values of the axion $\theta$. When the axion lies at its minimum at $\theta=0$, we see that the volume is stabilised at the Minkowski minimum. However, as we increase $\theta$ in units of the instanton coupling $a$, the volume barrier begins to shrink, and eventually disappears completely. At this point the volume will roll towards the AdS vacuum and any hope of exploiting the axion as a dynamical dark energy model is lost. 

These problems might have been anticipated in the racetrack scenario, as both the stabilisation of the volume and the dynamics of the corresponding axion rely on the {\it same} non-perturbative corrections to the superpotential. There was always a danger that the stabilisation would fail the moment the axion began to roll. As already pointed out, to proceed with a viable model of quintessence, we need to break the connection between the stabilisation of the volume and the dynamics of the would-be dark energy field, creating a hierarchy in mass between these two fields.

\subsection*{Saxion slow roll down a moderate slope, with a runaway or a Minkowski vacuum}

Let us now focus on a saxion runaway model, where the saxion is asymptotically rolling slowly down a moderate slope. 
At leading order, our example contains a non-supersymmetric Minkowski vacuum where one  of the saxion  directions is flat. The saxion runaway potential is then generated perturbatively, beyond leading order. However, since it is perturbative, it is not possible to generate a large enough hierarchy between the leading and subleading order terms to prevent the KL problem and destabilising the volume. Note that similar considerations would apply if additional subleading corrections were to generate a global non-supersymmetric Minkowski minimum, as opposed to a runaway. As shown in \cite{Hellerman:2001yi}, the case with a global supersymmetric Minkowski minimum is actually incompatible with slow roll down a moderate slope due to the stability condition on the form of the scalar potential.

Consider a fibred Calabi-Yau whose volume takes the form \cite{Cicoli:2011it,Cicoli:2018tcq}
\begin{equation}
\vo = \sqrt{\tau_1}\tau_2
\end{equation}
The saxion kinetic terms look like (we ignore the corresponding axions)
\begin{equation}
\mathcal{L}_{\rm kin}= \frac12 \left[(\partial \ln \tau_2)^2 +\frac12 (\partial \ln \tau_1)^2\right]
\end{equation}
and can be brought into canonical form by the following field redefinition 
\begin{equation}
\tau_1 = e^{\sqrt{\frac23} \chi + \frac{2}{\sqrt{3}}\phi}    \qquad\qquad \tau_2=e^{\sqrt{\frac23}\chi -\frac{1}{\sqrt{3}}\phi}
\end{equation}
Notice that $\chi$ corresponds to the volume mode $\vo$, and $\phi$ to the ratio $u=\tau_1/\tau_2$ since
\begin{equation}
\vo = \sqrt{\tau_1} \tau_2 = e^{\sqrt{\frac32}\chi} \qquad\qquad
u = \frac{\tau_1}{\tau_2} = e^{\sqrt{3}\phi}
\end{equation}
Let us consider an effective field theory defined by the following K\"ahler potential and superpotential
\begin{equation}
K = -2\ln\left(\vo+\frac{\xi}{2 g_s^{3/2}}-\gamma\sqrt{g_s} \ln \vo\right) - \frac{\tilde{C}}{\vo \sqrt{\tau_1}} \qquad\qquad W=W_0
\label{KandW}
\end{equation}
where $\xi$ controls $O(\alpha'^3)$ corrections, while $\gamma$ controls brane loop corrections at $O(\alpha'^3 g_s^2)$ \cite{Antoniadis:2018hqy}. For $\vo\gg 1$ these can naturally compete with the tree-level $O(\alpha'^3)$ term due to the $\ln\vo$ enhancement factor. The term proportional to $\tilde{C}$ represents $O(\alpha'^4 g_s^2)$ string loop corrections due to exchange of winding modes at the intersection of D7-branes. This contribution is subleading since it is suppressed by an additional power of $\alpha'$ with respect to the terms proportional to $\xi$ and $\gamma$. We do not include Kaluza-Klein loop correction since they would be suppressed with respect to winding contributions by an additional power of $g_s^2$, and, moreover, they could be absent by construction if all branes intersect each other. We also neglect higher derivative $F^4$ contributions to the scalar potential since they would arise with additional volume suppression factors.

The K\"ahler potential and superpotential in (\ref{KandW}) generate the following scalar potential
\begin{equation}
V = V_{\rm lead}(\vo) + V_{\rm sub}(\vo, u)
\end{equation}
where
\begin{equation}
V_{\rm lead}(\vo) = \frac{C_{\rm up}}{\vo^{8/3}} + \frac{3W_0^2}{4\vo^3} \left(-2\gamma\sqrt{g_s}\ln\vo+\frac{\xi}{g_s^{3/2}}\right)
\end{equation}
and
\be
V_{\rm sub}(\vo ,u)=   \frac{2 \tilde{C} W_0^2}{\vo^{10/3}} \frac{1}{u^{1/3}} 
\ee
Notice that in $V_{\rm lead}$ we included also a term proportional to $C_{\rm up}$ representing the positive contribution of a T-brane background \cite{Cicoli:2015ylx} which is a generic feature of type IIB compactifications with 3-form fluxes and magnetised D7-branes. In the limit where the supergravity approximation is under control, {\it i.e.} for $\vo\gg1$ and $g_s\ll 1$, $V_{\rm sub}$ is indeed subdominant with respect to $V_{\rm lead}$ since
\begin{equation}
\frac{V_{\rm sub}}{V_{\rm lead}} \sim \frac{g_s^{3/2}}{\vo^{1/3}} = \frac{g_s^2}{\vo_s^{1/3}} \ll 1
\end{equation}
where $\vo_s = \vo g_s^{3/2}$ denotes the string frame volume. Thus at leading order the potential features a flat direction parametrised by $u$. At this level of approximation, by a suitable tuning of $C_{\rm up}$, $V_{\rm lead}$ features a non-supersymmetric Minkowski minimum where the volume scales as (for $k=\xi/\gamma$)
\begin{equation}
\vo_{\rm min} \sim e^{\frac{k}{g_s^2}}\gg 1\qquad\text{for}\qquad g_s\ll 1
\label{min}
\end{equation}
The subleading contribution $V_{\rm sub}$ determines just a small shift of the volume minimum and generates a runaway for $u$ which, when written in terms of the canonically normalised field $\phi$, looks like
\begin{equation}
V_{\rm sub} (\phi)=   V_0\,e^{-\lambda\phi}\quad\text{with}\quad \lambda=\frac{1}{\sqrt{3}}\quad\text{and}\quad V_0 = \frac{2 \tilde{C} W_0^2}{\vo_{\rm min}^{10/3}}
\label{saxionrunaway}
\end{equation}
This gentle runaway could provide an interesting model of quintessence in agreement with the refined dS conjecture since $\lambda\simeq 0.577$ is of order unity and it marginally satisfies the bound $\lambda \le 0.6$ obtained in \cite{Agrawal:2018own}. However the requirement to avoid volume destabilisation due to the inflationary energy implies (similar considerations would apply also to the case where (\ref{saxionrunaway}) describes a quintessence potential with a global Minkowski minimum)
\begin{equation}
V_{\rm lead} = \left(\frac{V_{\rm lead}}{V_{\rm sub}}\right) V_{\rm sub} \sim \frac{\vo^{1/3}}{g_s^{3/2}} H_0^2\gtrsim H_{\rm inf}^2\quad \Leftrightarrow\quad \frac{\vo^{1/3}}{g_s^{3/2}} \gtrsim \left(\frac{H_{\rm inf}}{H_0}\right)^2 \gtrsim 10^{-36}\,,
\end{equation}
where we have used $H_{\rm inf}\gtrsim 10^{-42}\,M_p$ as the extreme lower bound on the Hubble scale during inflation to be compatible with a BBN reheating temperature of at least 1 MeV.
Using (\ref{min}) which implies $g_s \sim \left(\ln\vo\right)^{-1/2}$, this bound becomes
\begin{equation}
\vo^{1/3} \left(\ln\vo\right)^{3/4}\gtrsim 10^{36}\quad\Leftrightarrow\quad \vo \gtrsim 10^{103}
\end{equation}
This would yield a string scale $M_s$ well below the TeV scale and a gravitino mass $m_{3/2}$ well below the meV scale since
\be
M_s \simeq g_s^{1/4} \frac{M_p}{\sqrt{\vo}} \sim  \frac{M_p}{\left(\ln\vo\right)^{1/8}\sqrt{\vo}} \lesssim 10^{-52}\,M_p
\ee
\begin{equation}
m_{3/2} \simeq g_s^{1/2} \frac{M_p}{\vo} \sim  \frac{M_p}{\left(\ln\vo\right)^{1/4}\vo} \lesssim 10^{-104}\,M_p
\end{equation}
Similar considerations imply that the mass of the volume mode is also very suppressed with respect to the meV scale. Hence the hierarchy between $V_{\rm lead}$ and $V_{\rm sup}$ is not big enough to prevent the KL and light volume problems. The reason is that the effective shift symmetry for $u$ is already broken at perturbative level.

\section{Axion hilltop quintessence and initial conditions} 
\label{sec:hills}

As explained in the previous section, a viable quintessence model has to feature a leading order non-supersymmetric Minkowski vacuum with hierarchy, {\it i.e.} where, at leading order, the axion is a flat direction while the saxion (in particular, the volume mode) is heavy. The axionic flat direction is then lifted by subdominant instanton effects which can lead to axion hilltop quintessence. In this section we therefore focus on this model, providing first an explicit realisation in LVS string models, and then studying the issue of initial conditions.

\subsection{LVS axion hilltop quintessence}
\label{LVSaxion}

The simplest way to realise an axion hilltop quintessence model in type IIB string theory is through the simplest LVS scenario, with two K\"ahler moduli $T_b=\tau_b+{\rm i}\theta_b$ and $T_s=\tau_s+{\rm i}\theta_s$, where the `big' modulus $\tau_b$ turns out to be much larger than the `small' modulus $\tau_s$. The important point is that the scalar potential only  depends on the volume axion, $\theta_b$, at next to leading order, without affecting the stabilisation of the volume mode which occurs at leading order. As a result, the volume axion can potentially play the r\^ole of quintessence when it is rolling near the top of its potential, without having any of the adverse consequences we saw for the racetrack scenario. We should, however, be mindful of the fact that the simplest LVS setup leads to a non-supersymmetric AdS vacuum which needs to be uplifted to Minkowski by the inclusion of additional sources of energy like T-branes \cite{Cicoli:2015ylx}. Notice that these positive contributions to the scalar potential are generic features of consistent type IIB compactifications \cite{Cicoli:2011qg, Cicoli:2012vw, Cicoli:2013mpa, Cicoli:2013cha, Cicoli:2017shd} due to the presence of hidden sector stacks of D7-branes (induced by D7 tadpole cancellation), 2-form gauge fluxes (induced by Freed-Witten anomaly cancellation) and 3-form background fluxes (used to freeze the dilaton and the complex structure moduli). For further details and a comprehensive discussion of dS model building in string theory, see \cite{Cicoli:2018kdo, Danielsson:2018ztv}. 

The main ingredients of this model are $\mc{O}(\alpha'^3)$ contributions to the K\"ahler potential and non-perturbative corrections to the superpotential of the form
\be
K = -2\ln\left(\vo+\frac{\hat\xi}{2}\right) \qquad\qquad 
W =W_0 + A_s\,e^{-a_s T_s} + A_b\,e^{-a_b T_b}
\ee
where the internal geometry corresponds to a simple Swiss-cheese scenario with the volume given by $\vo=\tau_b^{3/2}-\tau_s^{3/2}$. The resulting potential looks like
\begin{eqnarray}
V&=& {4 A_b^2 a_b^2 \over 3\tau_b}\,e^{-2 a_b \tau_b}+{8 A_s^2 a_s^2\sqrt{\tau_s} \over3\tau_b^{3/2}}\,e^{-2a_s\tau_s}+{3|W_0|^2\hat{\xi}\over 4\tau_b^{9/2}}  - {4 A_b a_b \over \tau_b^2}|W_0|\,e^{-a_b\tau_b}  \cos (a_b\theta_b )  \nonumber
\\ &-& {4 A_s a_s\tau_s\over \tau_b^3}|W_0|\,e^{-a_s\tau_s}\cos (a_s\theta_s)
+{8 A_b A_s a_b a_s \tau_s\over \tau_b^2 }\,e^{-(a_b\tau_b+a_s\tau_s)}\cos\left(a_b\theta_b-a_s\theta_s\right), \nonumber
\end{eqnarray}
where we have used $\tau_b\gg \tau_s$, $a_b\tau_b\gg 1$, $a_s\tau_s \gg 1$ and we have assumed $W_0<0$ so that the axions are minimised at $\theta_b=\theta_s=0$. With the axions settled at their minima, we consider the stabilisation of $\tau_s$ and $\tau_b$. Dropping all terms suppressed by $e^{-a_b \tau_b}$ or more, one finds the well-known LVS results from the variation with respect to $\tau_s$ and $\tau_b$
\be
\vo\simeq \langle \tau_b\rangle ^{3/2} \simeq {3|W_0| \sqrt{\la\tau_s\ra}\over 4 A_s a_s}\, e^{a_s\la\tau_s\ra} \qquad\qquad
{\hat\xi\over 2}\simeq \la\tau_s\ra^{3/2} \simeq \left[{1\over a_s} \ln\left(\vo\over |W_0|\right)\right]^{3/2}
 \label{eq:stab_lvs}
\ee
The minimum is AdS, breaks supersymmetry and, to leading order, is given by
\be
\langle V \rangle \simeq  -{3\hat\xi |W_0|^2 \over 8 a_s \la\tau_s\ra \vo^3 }\,.
\ee
There are several sources of uplifting to Minkowski which can be expressed as 
\be
V_{\rm up}={\kappa \over \mathcal{V}^\alpha}\ , \label{eq:uplift_term}
\ee
where $\kappa$ is a positive coefficient and $0<\alpha<3$. For example,  $\alpha=8/3$ for T-branes while $\alpha=4/3$ for an anti D3-brane at the tip of a warped throat, although the particulars of the uplifting mechanism are unimportant for the discussion that follows. The uplift term modifies the second relation in \eqref{eq:stab_lvs} as
\begin{equation}
    {\hat{\xi}\over 2}=\la\tau_s\ra ^{3/2} - {2 \alpha \kappa\over 9 |W_0|^2} \mathcal{V}^{3-\alpha}\ , \label{eq:eqxi}
\end{equation}
and we will fix 
$\kappa$, to zeroth order in $e^{-a_b\tau_b}$, by demanding that the uplifted LVS vacuum is Minkowski, {\it i.e.}
\begin{eqnarray}
\langle V\rangle &=& -\left({|W_0|^2\la\tau_s\ra^{3/2} \over  \vo^3} - {\kappa \left(3-\alpha\right)\over 3\vo^\alpha}\right)=0 \nonumber \\
&\Rightarrow&\quad \kappa = {3|W_0|^2 {\la\tau_s\ra}^{3/2}\over \left(3-\alpha\right)\vo^{3-\alpha}} \ . \label{eq:eq3}
\end{eqnarray}
The equations \eqref{eq:stab_lvs}, \eqref{eq:eqxi} and \eqref{eq:eq3} form a system fixing $\left(\la\tau_b\ra,\la\tau_s\ra,\kappa\right)$ for a particular choice of $\left(\xi, W_0, A_s, A_b, a_s, a_b, \alpha\right)$. Once the Minkowski vacuum is fixed in this way, we focus on $\theta_b$ as a dark energy candidate. The hierarchy of scales between $e^{a_b \tau_b}$ and $e^{a_s \tau_s}$ guarantees that shifts in the $\theta_b$ direction do not destabilise the Minkowski vacuum. 

The uplift term can be further adjusted at $\mc{O}\left(e^{-a_b \tau_b}\right)$  to guarantee a Minkowski vacuum at $\tau_b=\la\tau_b\ra, \tau_s=\la\tau_s\ra$ and $\theta_b=\theta_s=0$. Releasing the volume axion, $\theta_b$, its dynamics is then described by the following dark energy potential to leading order
\begin{eqnarray}
V_{\rm DE}&=& \left[
{4 A_b a_b \over \la\tau_b\ra^2}|W_0|\,e^{-a_b\la\tau_b\ra}     -{8 A_b A_s a_b a_s \la\tau_s\ra \over \la\tau_b\ra^2}\, e^{-(a\la\tau_b\ra+a_s \la\tau_s\ra)}\right]\left(1-\cos (a_b\theta_b)\right) \nonumber \\
&\simeq &
{4 A_b a_b \over \la\tau_b\ra^2}|W_0|\,e^{-a_b\la\tau_b\ra}  \left(1-\cos (a_b\theta_b)\right) ,
\label{VDE}
\end{eqnarray}
where we explicitly see the Minkowski minimum at $\theta_b=0$. The maximum is located at $\theta_b=\pi/a$. 
From the form of the K\"ahler metric, the canonically normalised axion and the corresponding decay constant turn out to be
\be
\phi \simeq \sqrt{3\over 2} {\theta_b\over \la\tau_b\ra}\qquad \qquad f_a = \sqrt{\frac32}\frac{M_p}{a_b\la\tau_b\ra}\,,
\label{LVSfa}
\ee
so that (\ref{VDE}) can be rewritten in a more standard way as
\be
V_{\rm DE} = V_0 \left(1-\cos \frac{\phi}{f_a}\right) \qquad\text{where}\qquad V_0 \equiv {4 A_b a_b \over \la\tau_b\ra^2}|W_0|\,e^{-a_b\la\tau_b\ra} 
\label{VDELVS}
\ee
The $\eta$ parameter at the maximum of the dark energy potential, where $\theta_b=\pi/a_b$, becomes  
\be
\eta_{\rm hilltop}= \left.{V_{\text{DE}, \, \phi\phi}\over V_{\rm DE}} \right|_{\phi_{\rm max}}  =\left.\frac23 \la\tau_b\ra^2  {V_{\text{DE}, \, \theta_b\theta_b }\over V_{\rm DE}}\right|_{\theta_b=\pi/a_b} = -\frac13 a_b^2 \la\tau_b\ra^2 
\ee
To estimate this, notice that the value of the potential at the hilltop should be $\mc{O}(10^{-120})$ in Planck units to be compatible with dark energy at late times. This suggests $a_b \la\tau_b\ra\sim \mathcal{O}(100)$, and so $\eta_\text{hilltop}\sim- \mathcal{O}(3000)$. Clearly the curvature at the hilltop is compatible with the improved swampland bound of \eqref{dscon}. However, the large absolute value of the $\eta$ parameter requires a high degree of fine tuning of the initial conditions for $\theta_b$ if it is to give rise to a viable quintessence model as it rolls away from the maximum, as we will illustrate in Sec. \ref{sec:constraining}. 

\subsubsection*{Other approaches to axion hilltop quintessence}

Before discussing the issue of initial conditions, let us mention other two possible approaches which can lead to a viable quintessence models via axion hilltop:
\begin{enumerate}
\item \emph{Uplifted KKLT with an orientifold-odd axion}

The standard uplifted KKLT scenario with a single K\"ahler modulus $T=\tau + i \theta$ features a non-supersymmetric Minkowski vacuum with no flat direction. 
This cannot be used to drive quintessence since both $\tau$ and $\theta$ are lifted by the same non-perturbative effect. However, in the presence of an extra orientifold-odd modulus $G= c+ S b$ (where $S$ is the axio-dilaton), $b$ would also be lifted by the non-perturbative superpotential $e^{-a T}$ (with a minimum at $b=0$), while the axionic mode $c$ would remain flat. This axionic direction can, instead, be lifted at subleading order by the inclusion of fluxed E3-instanton corrections to $W$ of the form $e^{- a (T+S+i G)} \sim e^{-a T} e^{-a/g_s}$ for $b=0$ \cite{Cicoli:2021tzt}. Therefore, for $g_s\ll 1$, the scale of the potential for $c$ is exponentially suppressed with respect to the potential for $T$, providing a promising candidate for a viable quintessence model with decay constant $f_a\simeq  \sqrt{g_s/\tau}\,M_p \ll M_p$ for $\tau \gg 1$ \cite{Cicoli:2021gss}.

\item \emph{Non-geometric fluxes}

A second possibility is to consider the effect of non-geometric fluxes which extend the GVW superpotential to \cite{Shelton:2005cf, Benmachiche:2006df, Blaback:2013ht, Plauschinn:2018wbo}
\begin{equation}
W=P_1(U) + S P_2(U) + \sum_i T_i P^{(i)}_3(U)\ ,
\end{equation}
where $P^{(i)}_3$ are cubic polynomials of the complex structure sector $U$. Combined with the tree-level expression of the K\"ahler potential, the dependence of $W$ on $T$ generates no-scale breaking contributions to the scalar potential. 
With regards to the previous discussion, if we are able to stabilise all but one (axionic) modulus  at a non-supersymmetric Minkowski vacuum (or even a near Minkowski AdS minimum) at tree-level through an appropriate choice of fluxes, then the hierarchy between non-perturbative effects and tree-level would guarantee that lifting the leftover flat direction would not displace the heavy moduli from the tree-level minimum. Furthermore,  the leftover axion could be made parametrically light and may be used to drive quintessence.\footnote{Assuming that backreaction effects on the K\"ahler potential can be kept under control and the internal volume can be made large enough to trust the perturbative expansion.}
\end{enumerate}

\subsection{Constraining hilltops} 
\label{sec:constraining}

Hilltop models are classically unstable. The rate of the instability is controlled by the $\eta$ parameter, describing the rate of change of the gradient close to the maximum of the potential. The larger the $\eta$ parameter, in absolute value, the closer the field needs to start near the maximum in order to obtain the required period of acceleration. Fortunately, for quintessence, we only require one efolding of accelerated expansion (this is in contrast to early universe inflation which requires at least $50$). Nevertheless, in string theory, the $\eta$ parameter can sometimes be quite large forcing the field to start very close to the top of the hill. Classically, this is not problematic if one accepts the inevitable tuning of initial conditions, although as we will see later, quantum diffusion at early times can push the field away from the sweet spot, spoiling any realistic chance of late time quintessence. 

In this section we will derive the constraints on the parameters and the initial conditions of a generic model of late time acceleration. As explained in the previous sections, our main interest will be in axion hilltop models, although we will also generalise our analysis also to saxion hilltop models which we approximate in a neighbourhood of the maximum as an inverted quadratic.

\subsubsection*{Axion hilltop quintessence}

In the context of late time acceleration, axions are the prototype of thawing quintessence models \cite{Scherrer:2007pu, Dutta:2008qn}: models where the field is frozen due to Hubble friction until the very recent past. These models are known to be sensitive to the choice of initial conditions and relatively insensitive to the particular form of the potential. A generic axion potential has the usual trigonometric form
\begin{comment}
\be
V=V_0-\tilde{V_0}\cos\frac{\phi}{f_a}
\ee
\end{comment}
\be
V=V_0\left(1-\cos\frac{\phi}{f_a}\right)
\ee
where we have (for simplicity) assumed that the vacuum expectation value of the axion lies at vanishing potential, consistent with a Minkowski vacuum. This can lead to accelerated expansion in two distinct regimes $f_a>M_p$ and $f_a<M_p$.  For $f_a>M_p$  acceleration takes place in the concave region of the potential, whereas if $f_a<M_p$ it happens in the vicinity of the maximum. While models with super-Planckian decay constants are less sensitive to initial conditions, getting these large values for $f_a$ has proven challenging from a UV point of view due to the tension with the weak gravity conjecture \cite{Arkani-Hamed:2006emk} and with explicit computations \cite{Cicoli:2012sz, Cicoli:2021gss}. For example, the LVS axion model presented in Sec. \ref{LVSaxion} features for example a sub-Planckian decay constant since (\ref{LVSfa}) gives $f_a\simeq M_p/(a_b \la\tau_b\ra)\sim 0.01\, M_p$. 

\begin{figure}[h]
\begin{center}
\includegraphics[width=0.6\textwidth]{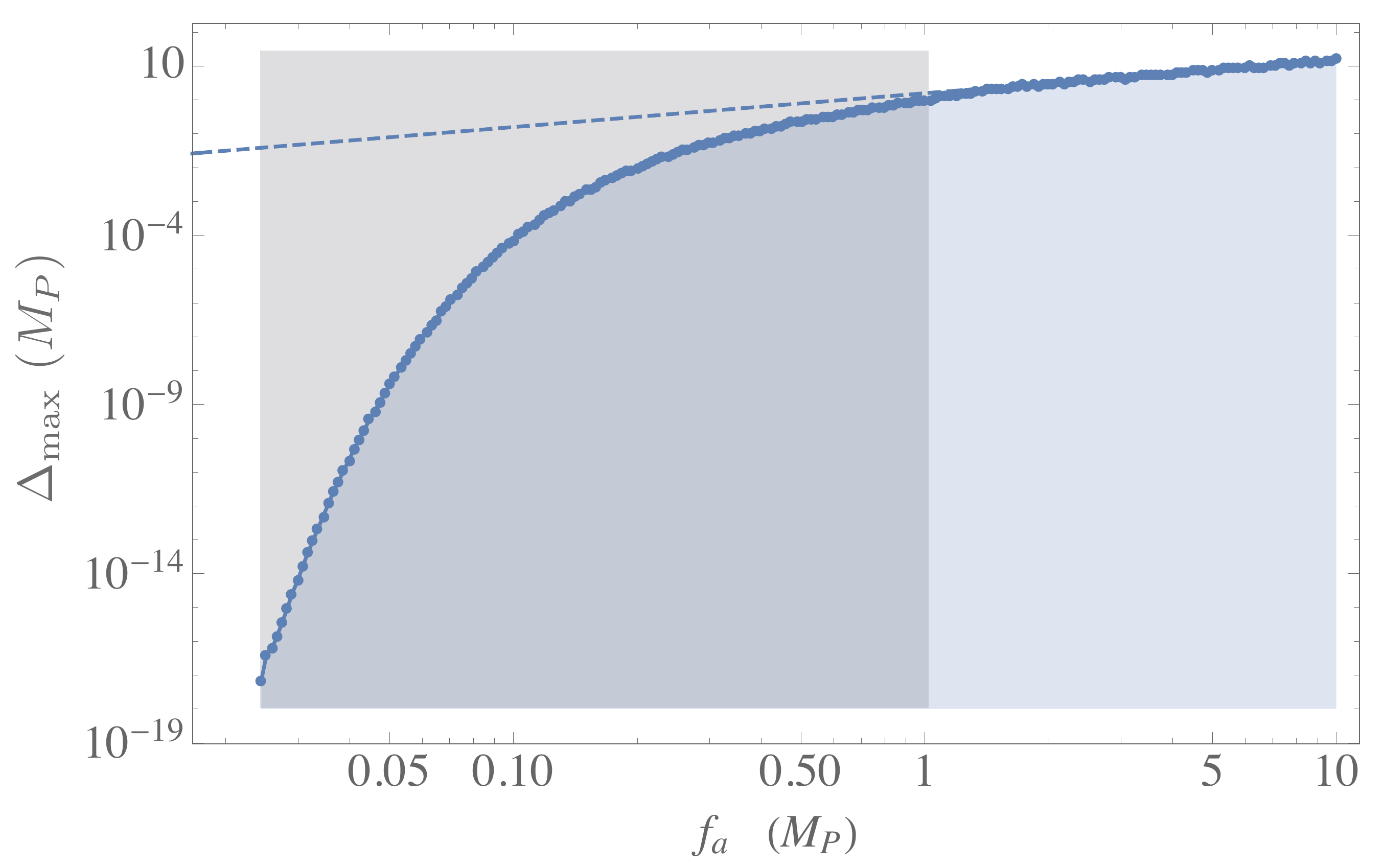}
\caption{Constraints on initial axion displacement from the maximum compatible with a viable quintessence hilltop model as a function of the decay constant $f_a$. For at least one e-fold of accelerated expansion, the initial value $\phi_\text{in}$ should  satisfy $|\phi_\text{in}-\phi_\text{max}|<\Delta_{\text{max}}$, where $\phi_\text{max}$ is the location of the maximum and $\Delta_{\text{max}}$ is given by the solid blue line. The dashed line shows the position of the inflection point $\phi=\pi f_a/2$. The blue shaded region corresponds to $f_a>M_p$ which is in tension with the weak gravity conjecture.}
\label{fig:f_vs_Phi0}
\end{center}
\end{figure}

Of course, axion hilltop quintessence can take place irrespective of the value of the decay constant, though it may require finely tuned initial conditions. In Fig. \ref{fig:f_vs_Phi0} we plot the deviation from the maximum as a function of the decay constant for a range of $f_a$ that is compatible with swampland constraints. We see that the range of $f_a$ that is more naturally achieved in UV constructions, $f_a<M_{\rm GUT}$, is also the one that suffers from an extreme sensitivity to the initial position of the field. In the region $f_a\in[0.02, 0.1]\,M_p$, the curve bounding the viable region can be approximated by
\be
\ln\Delta_{\text{max}} =c_0 +c_1 \ln f_a+c_2 (\ln f_a)^2\,,
\label{approximation}
\ee
where $\Delta_{\text{max}}$ denotes the maximum distance from the maximum compatible with late time acceleration, $c_0=-32.6$, $c_1=-28.977$ and $c_2=-8.2302$ .

\subsubsection*{Saxion hilltop quintessence}

For completeness, we now turn our attention to saxion models of quintessence, which, in the vicinity of the hilltop can approximated by an inverted quadratic
\be
V=V_0-\frac12\,m^2\, \phi^2.
\label{eq:V_HT}
\ee
It is useful to define the following parameter
\be
\eta_0=\frac{V_{\phi\phi}M_p^2}{V}\Big|_{\phi=0}=-\frac{\left(m M_p\right)^2}{V_0}
\ee
which describes the curvature of the scalar potential at the origin. Let us recall that the swampland conjecture (\ref{dscon}) requires $|\eta_0|\leq c'\sim\mc{O}(1)$. 

\begin{figure}[h]
\begin{center}
    \includegraphics[width= 0.5\textwidth]{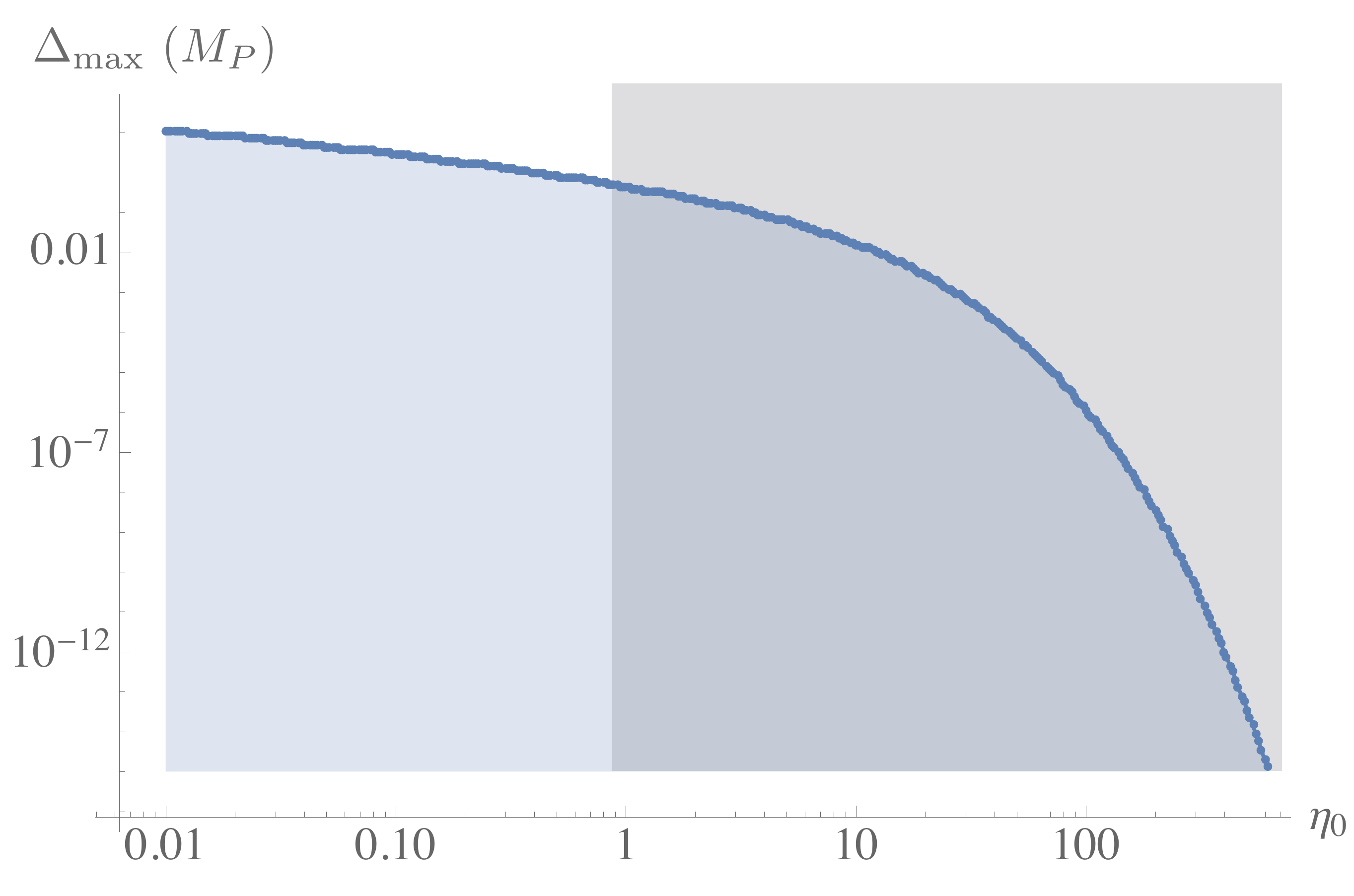}
\caption{Constraints on initial saxion displacement from the maximum compatible with a viable quintessence hilltop model as a function of $\eta_0$.  The grey region corresponds to $\eta_0>1$ which is tension with the refined swampland conjecture. The steeper the maximum (larger $\eta_0$), the more fine tuned the initial conditions (smaller $\Delta_{\text{max}}$).}
\label{fig:VKS}
\end{center}
\end{figure}

Hilltop models, no matter how steep the potential  (or how large $\eta_0$), \emph{classically} always lead to extended periods of accelerated expansion, given that at the maximum $\epsilon_V=0$. This would allow for a description of the late time acceleration, regardless of the swampland limits on the slope and curvature of the potential that can be attained within a UV complete framework. On the flip-side, this comes at the price of tuning the initial position of the quintessence field, $|\phi_{\rm in}|<\Delta_{\text{max}}$ -  the steeper $V$, the closer $\phi_{\rm in}$ needs to be to the maximum in order to have an extended period of acceleration. One would think that this problem is slightly less of an issue in quintessence models which require only $\mc{O}(1)$, instead of $\mc{O}(50)$, efoldings as in inflation. However for large $\eta_0$ the level of tuning is similar.  

In Fig. \ref{fig:VKS} we show the maximal allowed initial displacement from the maximum, $\Delta_{\text{max}}$, compatible with late time acceleration as a function of $\eta_0$ for quadratic quintessence models. Each point corresponds to numerical solutions that start in matter domination with zero initial velocity and different values for $\phi_{\rm in}$. We see that the steeper the potential the more tuned is the initial value of $\phi$. For $\eta_0\ge1$ (grey region) $\phi_{\rm in}$ must lie within a fraction of $M_p$ from the top of the hill. In the absence of a dynamical mechanism, such initial conditions look rather unnatural. The curve bounding the viable region can be approximated by
\be
\ln\Delta_{\text{max}} =c_1+c_2 \ \eta_0^{-p}\ ,
\ee
where $c_1=1.7$, $c_2=-2.1$ and $p=0.44$. 

Even if one is willing to accept this level of tuning in order to  describe  the observed accelerated expansion and be in agreement with putative bounds from a UV theory, one must ponder if quantum effects will spoil the required tuning of the initial conditions. We address this issue in Sec. \ref{sec:stochastic}.

\subsection{Stochastic effects and initial conditions}
\label{sec:stochastic}

In this section we investigate whether the judicious choices of initial conditions described in the previous section survive the unavoidable stochastic fluctuations in the early universe. For our purposes we model inflation as an exact dS background, and therefore fix $H_{\rm inf}$ to be constant. Generally speaking this approximation is adequate when $H_{\rm inf}$ varies slowly, like in plateau models, but needs to be refined in the context of monomial inflationary models as shown in \cite{Hardwick:2017fjo}. In the cases of interest, where the stochastic processes are diffusion dominated and the equilibrium distribution (if it exists) is of little relevance, the exact dS approximation gives an adequate description of the system.

The quintessence field $\phi$, due to the large hierarchy $V(\phi)\ll H_{\rm inf}^2 M_p^2$ behaves as a spectator during inflation  and is, to leading order in the slow roll expansion, described by the Langevin equation
\be
\frac{\partial \phi}{\partial N}=-\frac{V_\phi}{3 H_{\rm inf}^2}+\frac{H_{\rm inf}}{2\pi}\,\xi\ ,
\label{eq:Langevin}
\ee
where $N$ denotes the number of efoldings and $\xi(N)$ is a stochastic variable with unit variance $\langle \xi(N_1) \xi(N_2)\rangle=\delta(N_1-N_2)$ and zero mean $\langle \xi(N)\rangle=0$. The last term in (\ref{eq:Langevin}) describes the backreaction of the short-wavelength modes of $\phi$ onto the homogeneous mode and turns the deterministic slow roll evolution of $\phi$ into a stochastic process. The stochastic nature of \eqref{eq:Langevin} implies that the system can equivalently be described in terms of the Fokker-Planck equation for the probability density function $P(\phi,\phi_{\rm in},N)$
\be
\frac{\partial P}{\partial N}=\frac{1}{3 H_{\rm inf}^2}\frac{\partial }{\partial \phi}\left(V_\phi P\right)+ \frac{H_{\rm inf}^2}{8 \pi^2 }\, P_{\phi\phi}\ ,
\label{eq:Focker_Planck}
\ee 
where we take $\phi_{\rm in}$ to be fixed at the onset of inflation. Once the solution to \eqref{eq:Focker_Planck} is known, all relevant moments of the distribution can be computed:
\be
\langle\phi^n\rangle(N)=\int d\phi\  \phi^n  P(\phi,\phi_{\rm in},N)\ .
\ee
In what follows we will be interested in the  first two moments: the mean, $\langle\phi\rangle$, and the mean square, $\langle\phi^2\rangle$, which allow us to determine the variance of the distribution $\sqrt{\langle\phi\rangle^2-\langle\phi^2\rangle}$.

Given that the energy scale of quintessence is hierarchically smaller than the scale of inflation,  the quintessence field is classically frozen during the inflationary epoch. It is only expected to thaw once the background Hubble parameter drops to around $H\sim m_\phi$, which should happen during the matter phase, after the Big-Bang. This should hold true regardless of the shape of the quintessential potential.

The existence of a vast hierarchy between the value of the Hubble parameter today and during inflation, $H_0^2\simeq V_0/M_p^2\ll H_{\rm inf}^2$, implies that the quintessence field is a spectator during inflation, and that it is undergoing pure Brownian motion with \eqref{eq:Langevin} well approximated by
\be
\phi'=\frac{H_{\rm inf}}{2\pi}\,\xi \ ; 
\ee
or equivalently \eqref{eq:Focker_Planck}, by the one-dimensional heat equation
\be
\frac{\partial P}{\partial N}= \frac{H_{\rm inf}^2}{8 \pi^2}\, P_{\phi\phi}\,.
\label{eq:heat}
\ee
This dominance of the stochastic effects over the classical evolution has severe consequences for the retention of memory of the initial conditions for the quintessence field.
Exact solutions to \eqref{eq:heat} take the form (see e.g. \cite{Uhlenbeck:1930zz})
\be
P=\sqrt{\frac{2\pi}{N H_{\rm inf}^2}}\exp \left(-\frac{2 \pi^2 }{N}\frac{(\phi-\phi_{\rm in})^2}{H_{\rm inf}^2}\right)
\label{eq:heatKernel}
\ee
from which one can show that 
\be
\langle\phi\rangle =\phi_{\rm in}\ ,
\ee
{\it i.e.} the ensemble average is frozen at the specified initial value for the classical field $\phi_{\rm in}$, in accordance with the fact that classically the field is frozen by Hubble friction. One can also show that 
\be
\langle\phi^2\rangle =\left( \frac{H_{\rm inf}}{2\pi} \right)^{2} N +\phi_{\rm in}^2\ ,
\ee
which implies 
\be
\sqrt{\langle\phi^2\rangle-\langle\phi\rangle^2} =\frac{H_{\rm inf}}{2\pi} \sqrt{N}\ .
\label{eq:var_ax}
\ee
Therefore in one efolding of inflationary expansion, the spectator field will be kicked on average by $H_{\rm inf}/(2\pi)$. Depending on the sensitivity of a given hilltop to the choice of initial conditions, and on the exact value of $H_{\rm inf}$, these stochastic effects can push the field away from the top of the hill, and into a region where it cannot account for the observed present day accelerated expansion.

\begin{figure}[h]
\begin{center}
    \includegraphics[width= 0.5\textwidth]{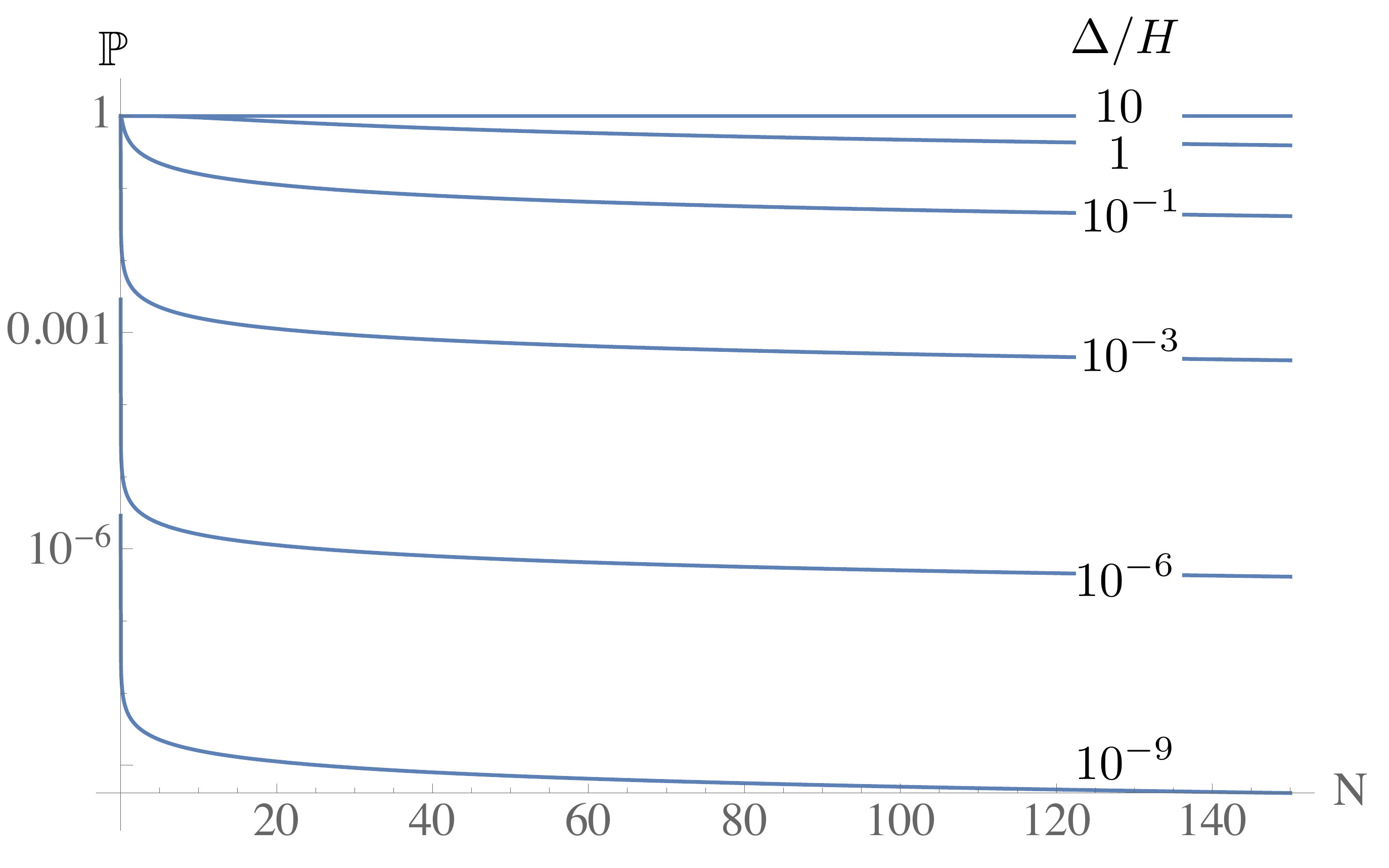}
\caption{Survival probability to remain within a distance $\Delta$ from the initial condition after $N$ efoldings of inflation, as a function of the ratio $\Delta/H_{\rm inf}$.}
\label{fig:Psurv}
\end{center}
\end{figure}

From \eqref{eq:heatKernel} we can compute the probability that a given choice of initial conditions survives the stochastic diffusion during a period of inflation. Setting $\phi_{\rm in}=0$ (assuming that this corresponds to the location of the maximum) and asking that after $N$ efoldings of inflation $\phi$ remains within a distance $\Delta$, we find that the survival probability is given by 
\be
\mathbb{P}(|\phi|\le \Delta)=\int_{-\Delta}^{\Delta} d\phi\,P= \text{erf}\left( \sqrt{\frac{2 \pi^2}{N}}\frac{\Delta}{H_{\rm inf}}\right)\ ,
\label{eq:survP}
\ee
where $\text{erf}$ is the complementary error function. For $N\gg 2\left(\frac{\pi \Delta}{H_{\rm inf}}\right)^2$ we can approximate \eqref{eq:survP} by
\be
\mathbb{P}(|\phi|\le \Delta)\simeq 2 \sqrt{\frac{2\pi}{N}}\frac{\Delta }{H_{\rm inf}}\,.
\label{important}
\ee
Once again we see that if $H_{\rm inf}\le \Delta$ the memory of the choice of initial conditions is preserved for a long period. In Fig. \ref{fig:Psurv} we plot the survival probability for various choices of $\Delta/H_{\rm inf}$.

\subsubsection*{Axion hilltop quintessence}

Stochastic effects on axionic hilltop quintessence models were studied in \cite{Kaloper:2005aj} and were pivotal in making the case for super-Planckian decay constants. Let us revisit these models in the light of the formalism reviewed above.

\begin{figure}[h]
\begin{center}
\includegraphics[width=0.6\textwidth]{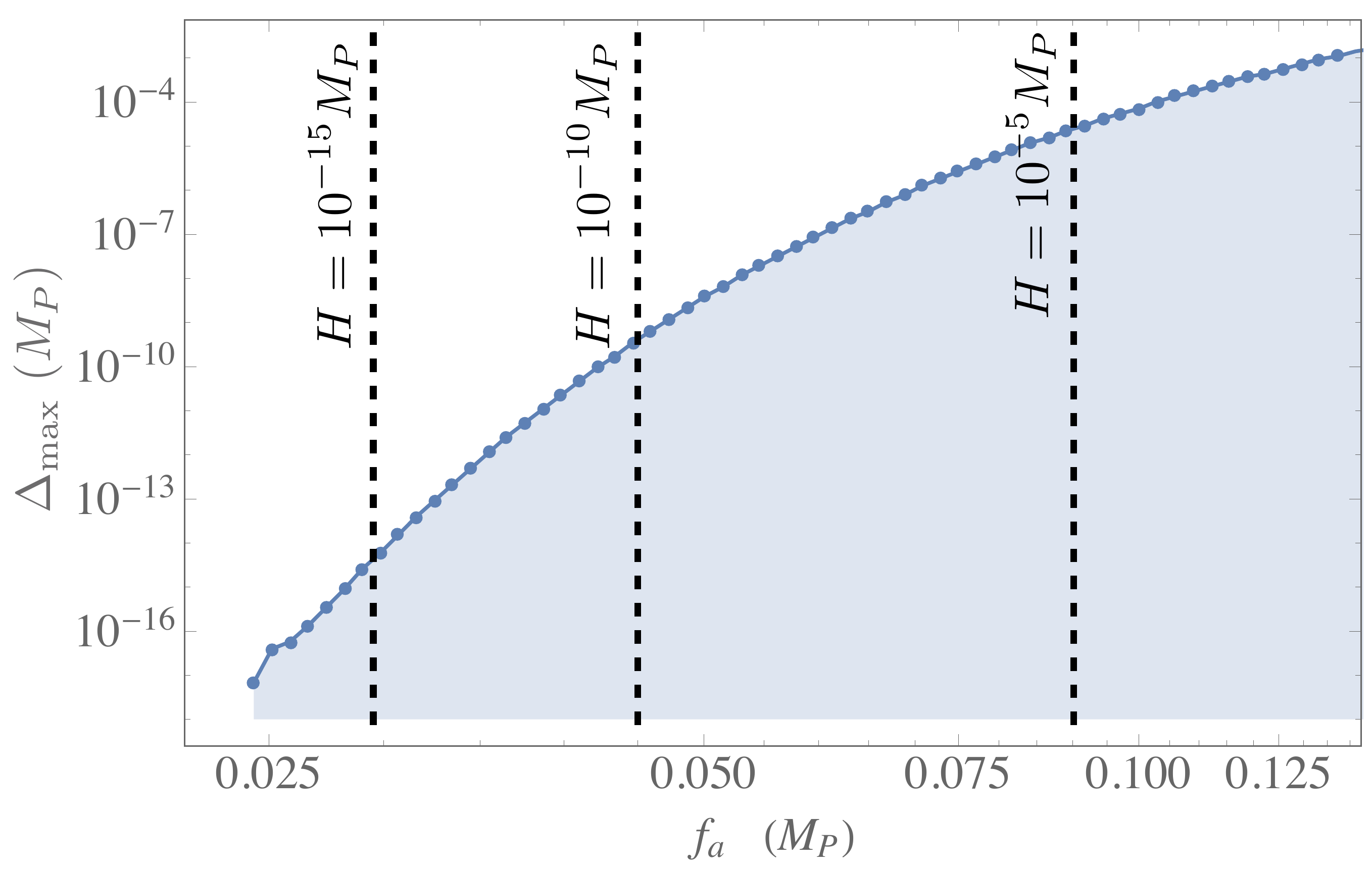}
\caption{Effect of quantum diffusion on the choice of initial conditions for axion hilltop quintessence models. Stochastic effects do not push the axion away from the hilltop region which yields late time acceleration only if $H_{\rm inf}\lesssim \Delta_{\text{max}}(f_a)$.}
\label{fig:f_vs_Phi0_ZOOM}
\end{center}
\end{figure}

Recall that current bounds on primordial tensor modes imply that, during inflation, $H_{\rm inf}\lesssim 10^{-5} M_p$ and that stochastic effects become relevant when the width of the region around the maximum of $V$ that gives rise to late time expansion is at, or below, the inflationary energy: $\Delta_{\text{max}}\lesssim H_{\rm inf}$. In Fig. \ref{fig:f_vs_Phi0_ZOOM} we zoom in on the low $f_a$ region of Fig. \ref{fig:f_vs_Phi0} and superimpose the constraints from $H_{\rm inf}$. We are led to the conclusion that axionic quintessence hilltop models with $f_a<0.1\,M_p$ are subject to stochastic fluctuations that (depending on the inflationary energy scale) may push $\phi>\Delta_{\text{max}}$ ruining the late time dynamics of those models. These estimates are in agreement with those of \cite{Kaloper:2005aj} and provide a worst case scenario. We note that a sharper statement can only be made with the knowledge of the exact energy scale of inflation.

\begin{figure}[!t]
\begin{center}
\includegraphics[width=0.47\textwidth]{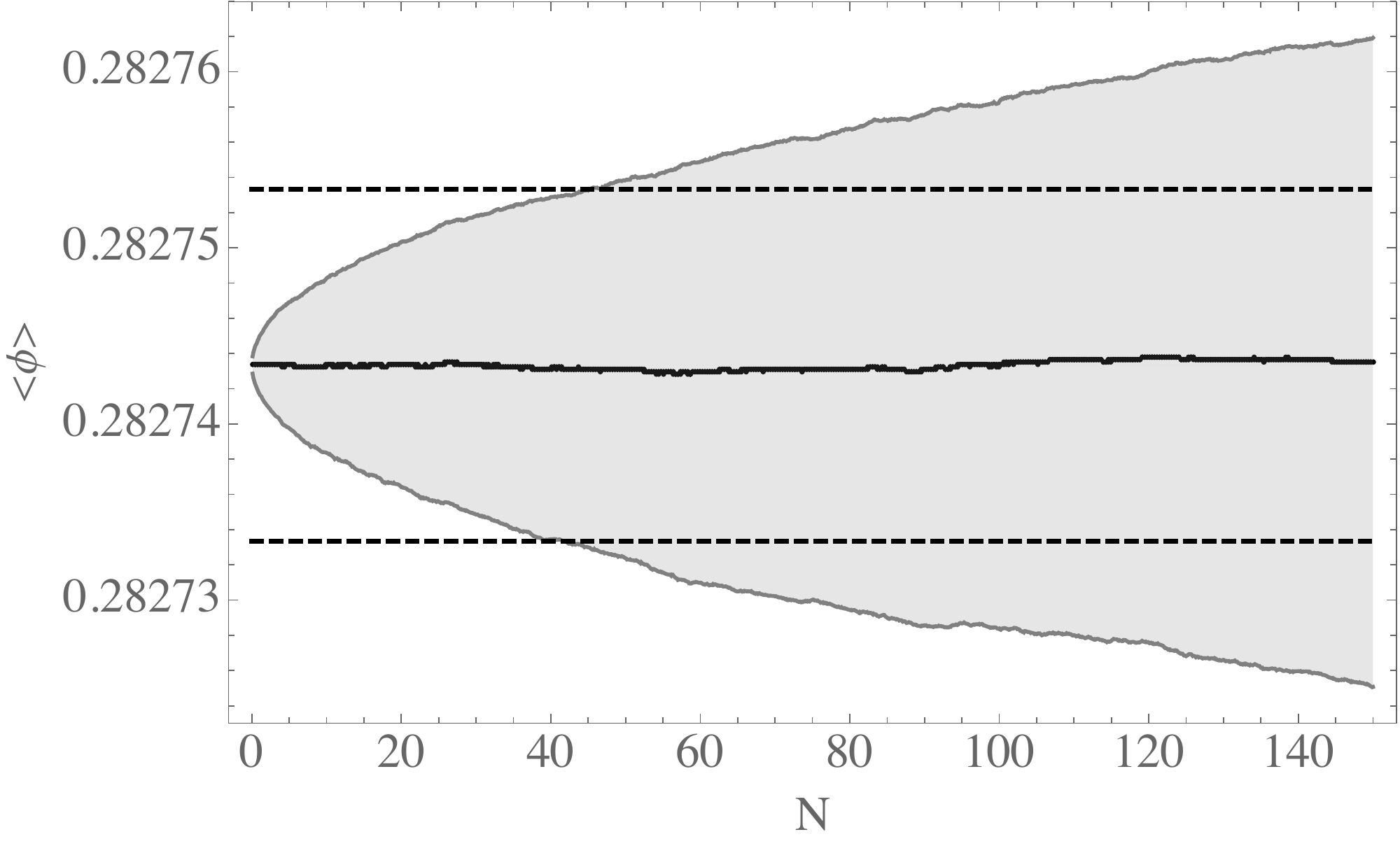}\hspace{0.5cm }\includegraphics[width=0.47\textwidth]{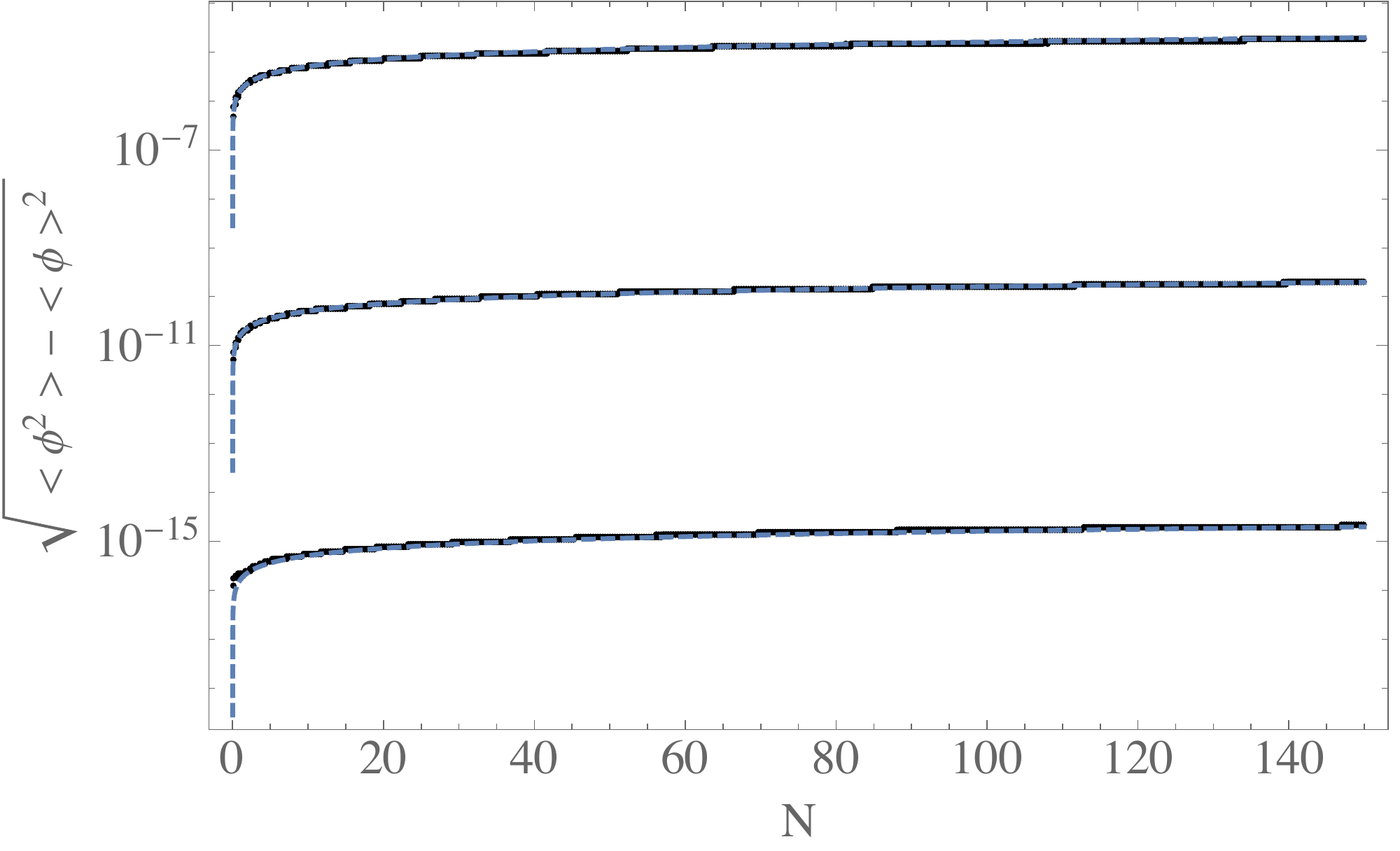}
\caption{Stochastic evolution for an axionic hilltop spectator with $f_a=0.09\, M_p$, $V_0=10^{-120} M_p^4$ and $\phi_{\rm in}=\pi f_a$ from $1000$ numerical solutions of the Langevin equation. Left panel: $H_{\rm inf}=10^{-5} M_p\simeq \Delta_{\text{max}}$, the grey region corresponds to the $1 \sigma$ band and the grey dashed lines denote the interval around the hilltop where the field should find itself in the matter phase to be able to drive quintessence. Right panel: standard deviation for (top to bottom) $H_{\rm inf}=\{ 10^{-5}, 10^{-10}, 10^{-15}\} \ M_p$, the black line corresponds to the $1000$ realisations of the Langevin equation while the dashed blue line corresponds to the analytical solution of \eqref{eq:var_ax}.}
\label{fig:P_FP_ax}
\end{center}
\end{figure}

In Fig. \ref{fig:P_FP_ax} we plot the solutions to the Langevin and Fokker-Planck equations for an axionic spectator with $H_{\rm inf}\simeq \Delta_{\text{max}}$, where it is evident that stochastic effects lead to a loss of memory of initial conditions after $N>\mc{O}(10)$ efoldings of inflation. 

The situation is even worse for $H_{\rm inf}>\Delta_{\text{max}}$ since the probability of lying within a distance $\Delta_{\text{max}}$ away from the maximum after $N\simeq 50-60$ efoldings of inflation turns out to be extremely small, as can be seen from (\ref{important}) with $\Delta = \Delta_{\text{max}}(f_a)$:
\be
\mathbb{P}(|\phi|\le \Delta_{\text{max}}(f_a))\simeq 2 \sqrt{\frac{2\pi}{N}}\frac{\Delta_{\text{max}}(f_a)}{H_{\rm inf}}\,,
\label{important2}
\ee
where $\Delta_{\text{max}}$ depends on the axion decay constant $f_a$ (for $f_a\in[0.02, 0.1]\,M_p$, $\Delta_{\text{max}}$ is very well approximated by (\ref{approximation})). For example, in the explicit LVS axion model of Sec. \ref{LVSaxion}, equations (\ref{LVSfa}) and (\ref{approximation}) give a decay constant $f_a \simeq 0.02\,M_p$ and a maximum displacement, $\Delta_{\text{max}}\simeq 2.4\times 10^{-20}\,M_p$. Choosing the largest value of $H_{\rm inf}$ compatible with the lack of observation of primordial gravity waves, $H_{\rm inf}\simeq 2\times 10^{-5}\,M_p$, and $N\simeq 50$ (\ref{important2}) would give $\mathbb{P}(|\phi|\le \Delta_{\text{max}})\simeq 10^{-15}$. Clearly smaller values of $H_{\rm inf}$ would give a smaller survival probability. Notice from Fig. \ref{fig:f_vs_Phi0} that $\Delta_{\text{max}}$ drops very quickly for smaller values of $f_a$, reducing the survival probability even further. 

We therefore conclude that a viable axion hilltop quintessence model requires two crucial conditions: ($i$) a tuning of initial conditions close to the maximum which becomes more severe for smaller values of $f_a$; and ($ii$) $H_{\rm inf}\lesssim \Delta_{\text{max}}$, else  stochastic effects will very quickly push the field away from the hilltop region compatible with a late time period of accelerated expansion. This second constraint turns out to be very strong since explicit computations of axion decay constants from string theory typically yield $f_a\lesssim 0.02\,M_p$ in the regime where the effective field theory is under control. For these small values of $f_a$, the maximum displacement is bounded as $\Delta_{\text{max}}\lesssim 10^{-20}\, M_p$\footnote{Numerically, we did not consider decay constants $f_a<0.02\,M_p$ since the high degree of tuning of the initial conditions rapidly brings about numerical precision issues. This prevent us from explicitly determining $\Delta_{\text{max}}$ for such low values of $f_a$, although the result can be obtained by extrapolating the validity of the formula \eqref{eq:heat}. In any event, the precise estimate is not really needed for drawing our general conclusions.}. Inserting this into the second constraint above, we see that we require a very low scale of inflation, $H_{\rm inf} \lesssim 10^{-20}\, M_p$.
When combined with the observed amplitude of scalar perturbations $A_s$, this bound on $H_{\rm inf}$ can then be translated into a severe upper bound on the inflationary slow roll parameter
\be
\epsilon_V = \frac{1}{8\pi^2 A_s}\left(\frac{H_{\rm inf}}{M_p}\right)^2 \lesssim 10^{-35}\,.
\ee
This is in strong tension with the dS swampland conjecture (\ref{dscon}) which requires $\mc{O}(1)$ values of $\epsilon_V$ (unless inflation is also realised extremely close to a maximum). Thus we conclude that axion hilltop quintessence would either be in the swampland or, if we ignore the refined dS swampland conjecture, would require a high tuning of initial conditions combined with a Hubble scale during inflation below $\mc{O}(1-10)$ MeV. Because these models are so contrived, we expect that dynamical dark energy, if supported by data, will have to be driven by a different mechanism, probably along the lines of axion alignment \cite{Kim:2004rp}.

\subsubsection*{Saxion hilltop quintessence}

The effects of diffusion in saxion hilltop models are qualitatively similar to those of axionic models. They will constrain the steeper hilltops, as these are the ones where the initial conditions are more severely tuned. From Fig. \ref{fig:VKS} we see that quadratic hilltops with $\eta_0\gtrsim 70$ require $\Delta_{\text{max}}\lesssim 10^{-5} M_p$ rendering them potentially vulnerable to diffusion effects during inflation, as CMB observations imply $H_{\rm inf}\lesssim 10^{-5} M_p$. Notice that the results for the effects of quantum diffusion obtained for the axion case can  also be used for saxion hilltop quintessence as long as we identify $|\eta_0| = \left(M_p/f_a\right)^2$.

\section{Conclusions}
\label{Conclusions}

With compelling observational evidence for dark energy \cite{Perlmutter:1998np,SupernovaSearchTeam:1998fmf, Aghanim:2018eyx},  we cannot avoid the question of its microscopic origin. But should we be looking for a cosmological constant or quintessence? If the latter, then is it driven by a scalar or pseudo-scalar, on a shallow potential or at a hilltop? At present all options are observationally viable, but we can also ask which is easiest to build into a fundamental theory. 

In this paper, we have outlined several challenges facing string theory models of quintessence focusing on effective field theories where perturbation theory is under numerical control - i.e. where the dilaton, ${\rm Re}(S)\gg 1$, and the volume mode, $\vo\gg 1$, are large enough to trust both the string loop and the $\alpha'$ expansion. This is arguably the most interesting region of moduli space since deep in the bulk, where ${\rm Re}(S)\sim \vo \sim \mc{O}(1)$, one would need a full knowledge of the whole quantum theory, while at boundary of the moduli space, where ${\rm Re}(S)\to\infty$ and $\vo \to \infty$, there is strong evidence indicating the absence of both dS vacua \cite{Ooguri:2018wrx} and a viable quintessence dynamics \cite{us, Hertzberg:2007wc, Garg:2018zdg, ValeixoBento:2020ujr}. 

Some of the problems of dynamical dark energy models are shared with the pure cosmological constant (like the smallness of $H_0$), while others are particular to quintessence (like constraints from fifth forces, tuning of initial conditions and radiative stability of the mass of the quintessence field). Here we were particularly concerned with  the destabilisation of the volume modulus  during inflation (the KL problem applied to quintessence) and generating the large hierarchy between the scale of the universe today and during inflation.  We have argued that the need to overcome all of these challenges has singled out a preferred model building scenario for dynamical dark energy. 
%We have argued that the need to avoid these problems - most notably,  the destabilisation of the volume during inflation (the KL problem applied to quintessence) and the large hierarchy between $H_0$ and $H_{\rm inf}$, singles out a preferred model building scenario for dynamical dark energy. 
The leading order contributions to the scalar potential should yield a vacuum with the following properties:
\begin{enumerate}
\item it should admit a flat direction in order to decouple the dark energy scale from the inflationary scale;
\item this flat direction should be axionic. This is because saxions are already lifted  at perturbative level without being able to generate the required hierarchy between $H_0$ and $H_{\rm inf}$. Axions, in contrast, develop a potential via highly-suppressed non-perturbative effects;
\item it should be (nearly) Minkowski since otherwise subleading corrections would not be able to push it up to the positive energies required to drive an epoch of accelerated expansion; 
\item it should break supersymmetry in order to decouple the gravitino mass from the dark energy scale. 
\end{enumerate}

It is interesting to combine these results with the swampland dS conjecture that would rule out dS vacua. From a model-building perspective, dS vacua look qualitatively the same as  non-supersymmetric Minkowski, leaving dynamical dark energy as the only explanation for the present acceleration of the universe. However we have found that it is extremely hard to realise a working model of quintessence in any scenario which would be in agreement with the swampland dS conjecture, like moderately sloped runaways, or supersymmetric AdS or Minkowski vacua. This tension raises some doubts on the validity of the swampland dS conjecture since it would imply that quantum gravity is in contradiction with observations. 

At this point it is tempting to favour the humble cosmological constant as the simplest empirical model of dark energy: it fits the available data and avoids the additional complications associated with quintessence. However, it is important to note that quintessence can open up opportunities to solve other cosmological problems.  For example, in \cite{Cunillera:2021izz}, it was shown how dynamical models of quintessence in string theory may shed new light on the cosmological coincidence problem \cite{whynow,Zlatev:1998tr,Velten:2014nra}. An evolving scalar on cosmological scales may also allow for self adjustment mechanisms to address the naturalness problems associated with vacuum energy (see \cite{Weinberg:1987dv, Niedermann:2017cel} for relevant no go theorems, and \cite{Copeland:2021czt} for a recent way around them). But perhaps most importantly,  future observations may rule out the cosmological constant as the driver of late time acceleration. 

If this were indeed the case, our analysis provides guidance for successful quintessence model building in string theory. In fact, we studied axion hilltop quintessence  in detail since vanilla string compactifications lead to axion decay constants at least two orders of magnitude below the Planck scale. We found that hilltop models are rather contrived since, even if the initial conditions are tuned very close to the maximum, quantum diffusion effects during inflation would kick the quintessence field away from the accelerating region close to the maximum, unless the Hubble scale during inflation is extremely low, $H_{\rm inf}\lesssim \mc{O}(1-10)$ MeV. Of course, one could envisage a scenario where a suitable coupling between the inflaton and the quintessence field makes the latter heavy during inflation, thereby suppressing any stochastic effect. However, after the end of inflation, the inflaton would typically settle down at the minimum of its potential, reaching its present day value. Quantum diffusion would then still play an important r\^ole in the reheating phase  and after, implying that the results of Sec. \ref{sec:stochastic} would still hold qualitatively after inflation with $H_\text{inf}$ replaced by the Hubble scale during a given epoch. We conclude that a more promising avenue to build a working model of dynamical dark energy is to rely on alignment mechanisms to obtain an effective axion decay constant which is trans-Planckian \cite{Kim:2004rp}.

As stated earlier, for dynamical dark energy, we first seek a scenario where the volume is stabilised at leading order to a vacuum that is uplifted to Minkowski. Non-perturbative corrections can then be exploited to drive dark energy at the correct scale. Although it might seem a little uneconomical to uplift and then do quintessence, if dark energy turns out to be dynamical both steps may be necessary to explain the present state of the universe in the context of string compactifications. 

In truth, both the cosmological constant and quintessence face formidable challenges from the perspective of consistent model building in string theory, while remaining perfectly compatible with observational constraints. It behooves us to  better understand the limitations imposed by perturbative string theory in both cases. Indeed, does a microscopic understanding of dark energy require input from non-perturbative strings, through string field theory, or M-theory?  Since this may be a question of properly understanding the vacuum structure of the theory, this seems like a reasonable possibility.

\section*{Acknowledgements}

We would like to thank Arthur Hebecker, Susha Parameswaran, Fernando Quevedo and Ivonne Zavala for useful discussions. AP was supported by an STFC consolidated grant number ST/T000732/1 and FC by a University of Nottingham studentship.

\end{document}